\documentclass[preprint]{aastex}

\begin{document}
    
\title{Mapping the Galactic Halo II: Photometric Survey}

\author{R.C. Dohm-Palmer}
\affil{Astronomy Department, University of Michigan, Ann Arbor, 
MI 48109\\ and\\ Steward Observatory, University of Arizona, Tucson, AZ 
85721}
\email{rdpalmer@astro.lsa.umich.edu}

\author{Mario Mateo}
\affil{Astronomy Department, University of Michigan, Ann Arbor, 
MI 48109}
\email{mateo@astro.lsa.umich.edu}

\author{E. Olszewski}
\affil{Steward Observatory, University of Arizona, Tucson, AZ 85721}
\email{edo@as.arizona.edu}

\author{H. Morrison \altaffilmark{2,3}}
\affil{Astronomy Department, Case Western Reserve University, 
Cleveland, OH 44106}
\email{heather@vegemite.astr.cwru.edu}
\altaffiltext{2}{Cottrell Scholar of Research Corporation and NSF 
CAREER fellow}
\altaffiltext{3}{and Department of Physics}

\author{Paul Harding}
\affil{Steward Observatory, University of Arizona, Tucson, AZ 85721\\ 
and\\ Astronomy Department, Case Western Reserve University, 
Cleveland, OH 44106}
\email{harding@billabong.astr.cwru.edu}

\author{Kenneth C. Freeman and John Norris}
\affil{Mount Stromlo and Siding Spring Observatories ANU, Private 
Bag, Weston Creek PO, 2611 Canberra, ACT Australia}
\email{kcf@mso.anu.edu.au, jen@mso.anu.edu.au}

\begin{abstract} 

We present imaging results from a high Galactic latitude survey 
designed to examine the structure of the Galactic halo.  The objective 
of the survey is to identify candidate halo stars which can be 
observed spectroscopically to obtain radial velocities and confirm 
halo membership.  The Washington filter system is used for its ability 
to distinguish between dwarfs and giants, as well as provide a 
metallicity indicator.  Our most successful imaging run used the BTC 
camera on the CTIO 4m telescope in April 1999.  Photometric conditions 
during these observations provided superb photometry, with average 
errors for a star at $M=18.5$ of 0.009, 0.008, 0.011, and 0.009 for 
$C$, $M$, $DDO51$, and $T2$ respectively.  These data are available 
with the electronic version of this paper, as well as through ADC 
(http://adc.gsfc.nasa.gov/).  We use these data as a template to 
describe the details of our photometric reduction process.  It is 
designed to perform CCD reductions and stellar photometry 
automatically during the observation run without the aid of external 
packages, such as IRAF and IDL. We describe necessary deviations from 
this procedure for other instruments used in the survey up to June 
2000.  Preliminary results from spectroscopic observations indicate a 
97\% efficiency in eliminating normal dwarfs from halo giant 
candidates for $M<18.5$.  Unfortunately, low-metallicity subdwarfs 
cannot be photometrically distinguished from giants using the 
Washington filters.  These major contaminates unavoidably reduced the 
overall giant identification efficiency to 66\% for $M<18.5$.  Our 
improved knowledge of these stars will increase this efficiency for 
future spectroscopic observations.
\end{abstract}


\section{Introduction}

A revolution is underway in our understanding of how galaxies form: 
The idea of a monolithic collapse of a single system at earliest times 
\citep{egg62} has been challenged by observations at high redshift and 
by increasingly sophisticated models of the evolution of structures 
within the basic framework of the Hot Big Bang model \citep{ste94}.  
These findings imply that we should see clear evidence of hierarchical 
formation processes within nearby galaxies.  Indeed, some of the 
earliest work that suggested complex and possibly hierarchical 
evolutionary histories for galaxies such as ours came from studies of 
the stellar populations of the outer halo of the Milky Way 
\citep{sea78}.  Studies of nearby and high-redshift systems represent 
complementary approaches.  No viable model of galaxy formation can be 
considered successful unless the evidence from both sources can be 
understood within a single framework.

The fossil record of the local galaxian population remains sparsely 
sampled.  The halo is the most likely component to retain information 
of the Galaxy's formation history.  Apart from the isolated (and 
possibly peculiar) star clusters and dwarf galaxies, much of our 
knowledge of the halo derives from a relatively small number of nearby 
stars `passing through' the solar neighborhood.  Surveys that extend a 
few kpc from the Sun isolate halo stars more effectively, but even 
so, contamination by thick disk stars and background galaxies can be 
significant.  The remoteness of the Magellanic Clouds and M31 make 
direct efforts to identify substructure in their halos extremely 
challenging.

Despite the difficulties, recent studies of our Galaxy have begun to 
reveal tantalizing evidence of possible substructure in the halo 
\citep{maj92, cot93, arn92, hmi99, yan00}.  More spectacularly, the 
discovery of the SGR dwarf galaxy \citep{iba94} provides indisputable 
evidence that our Galaxy's halo is accreting stars and clusters that 
formed and evolved in an independent galaxy.  The question has now 
shifted from `Did accretion play any role in the formation of the 
Galaxy?'  to `How much of the Galaxy formed from accretion of 
hierarchical fragments?'

This paper continues a series that describes our efforts to address 
this second question by identifying substructure directly within the 
Galactic halo.  An overview of this survey was presented in 
\citet[Paper I]{mor00}.  To summarize, we are carrying out a 
large-scale study of high-latitude fields optimized to identify 
distant halo stars as potential tracers of substructure in the outer 
Galaxy.  Our observational approach is sensitive to a variety of halo 
stellar inhabitants, including (a) distant halo blue horizontal branch 
stars, (b) blue metal-poor stars \citep[BMP]{pre94}; observationally, 
these are field analogs of blue stragglers found in star clusters 
\citep{bai95}, (c) halo turnoff stars, and (d) red giants.  The bluer 
stars can be identified photometrically with relative ease using any 
number of combinations of broad-band colors.  The Sloan Digital Sky 
Survey \citep{yan00} and our survey (Paper I) have already proven very 
effective at doing so. These can be considered wide-field extensions 
of the surveys of Hawkins \citep{nor91}.

Identification of the red giants, however, demands more precise and 
specialized photometric observations and added care to identify and 
remove contaminants.  But this effort does have important rewards: red 
giants are the most luminous tracers found in significant numbers in 
the halo and hence allow us to probe to the greatest depth.  
Conceivably, luminous giants can be identified at distances as large 
as 200 kpc in our survey, though the giant-branch luminosity function 
ensures that the bulk of the stars we find will be between 20-100 kpc 
distant.  Certain important biases that are inherent in the studies of 
horizontal branch stars \citep{kin94} are also avoided by isolating 
red giants whose properties are well understood as a function of age 
and metallicity.  In addition, studies of the chemical evolution of 
the outer halo will be possible via detailed abundance analysis of 
halo giants using 10m-class telescopes.  These studies are much more 
difficult for the hotter horizontal-branch stars.

The purpose of this paper is twofold: to describe the procedures we 
have developed to carry out the photometric observations of our 
survey, and to publish the photometry of photometrically calibrated 
fields.  Our approach employs the modified Washington photometric 
system \citep{can76, gei84} to provide indices that can be used to 
isolate the four classes of stars described above.  We use four 
filters in our survey to produce indices sensitive to temperature 
($M-T2$), surface gravity/luminosity ($M-51$), and metallicity 
($C-M$).  These indices provide the means of selecting halo giants 
from the blanket of thin and thick disk stars, and unresolved 
galaxies, the sum of which dominate the point sources in any halo 
field.  It should be noted that the luminosity discrimination of the 
51 filter is based on the strength of Mg hydride band and Mg b lines.  
Thus, at low metallicities the ability to photometrically distinguish 
subdwarfs from giants disappears \citep{mor00b}.

Our observational procedures were first developed for use at the 
Burrell Schmidt \footnote{Observations were made with the Burrell 
Schmidt of the Warner and Swassey Observatory, Case Western Reserve 
University} and CTIO Schmidt telescopes.  The basic approach was 
subsequently carried over to the NOAO 4m telescopes where our survey 
efficiency is much higher.  The first of these runs employed the BTC 
camera on the CTIO 4m telescope in April 1999; it has proven to be the 
most successful run to date.

A subset of photometric red giant and horizontal branch candidates
produced from these data sets have been observed spectroscopically.
These observations provide a test for our photometric selection
techniques and will be briefly discussed here.  More details on the
spectroscopy can be found in Paper I and in future papers of this
series.  Two important conclusions from these spectroscopic
observations are worth stating here.  

First, selection of distant red giants {\it demands} high-quality 
photometry to not only define the luminosity index ($M-51$), but also 
to minimize contamination from the many stellar and galaxian 
contaminants found in all fields.  When foreground dwarfs outnumber 
the rare giants by orders of magnitude, even three- or four-sigma 
errors can produce red-giant imposters.  Our photometric procedures 
represent a balance between the need for precise photometric indices, 
and the need for an efficient, deep, survey of a significant fraction 
of the halo.  Second, even with perfect photometric observations and a 
highly restrictive set of selection criteria, a photometric sample of 
distant red giants will be contaminated by metal-poor subdwarfs within 
the halo.  Spectroscopic confirmation is necessary to identify and 
remove these relatively nearby stars from any deep halo survey such as 
ours \citep{mor00b}.

The price of an efficient survey employing mosaics of CCD cameras is a 
very high data rate; on good nights we obtain 10-15 Gbytes of raw 
imaging data with the NOAO Mosaic imagers.  These data must be reduced 
quickly to avoid serious backlogs and, more importantly, to promptly 
identify halo candidates that can be subsequently observed 
spectroscopically.  Our imaging and spectroscopic observations are 
often scheduled within a few weeks of each other to be able to survey 
the same high-latitude regions in a given season.  Throughout this 
survey -- which has seen an increase in the data rate of over a factor 
of 50 in 1.5 years -- we have aimed to carry out the full processing 
and photometric reductions of our imaging data in near real-time at 
the telescope.  We have developed software that makes it possible to 
achieve this goal without using any additional programs or packages 
such as IRAF or IDL and the associated overheads and costs.  This 
process is robust and flexible enough to handle the inevitable problem 
images within a large data set.

The following section summarizes the observations, and describes the 
fields that have been observed as of June 2000.  Section \ref{pipe} 
describes the automated pipeline for processing images and measuring 
photometry and, ultimately, astrometry of sufficient precision to 
support single-slit or multi-object spectroscopy.  Section 
\ref{results} presents the photometric results, emphasizing the 
precision that we achieve and the efficiency of our photometric 
selection procedure.

\section{Observation Summary}

\subsection{Field Selection}
\label{fields}

The goal of this survey is to identify distant halo stars.  It is 
advantageous to avoid the confusion of fields near the Galactic plane 
where the fraction of halo stars is extremely small.  Thus, with a few 
exceptions, all fields were chosen to have a Galactic latitude of $|b| 
\ga 25$.

Furthermore, we wish to avoid highly reddened areas to avoid the 
possibility of differential reddening across a field, and to prevent 
distant halo stars being extincted below detection limits. All 
potential fields are checked on the reddening maps of \citet{sch98}. 
Fields with high extinction, or spatially varying reddening are 
removed. With 1 exception, all fields have $E(B-V) < 0.13$. 

Finally, we wish to avoid placing very bright stars in the images.  
These will saturate the surrounding region of the CCD chip, and add 
diffraction spikes that can confuse photometric reductions.  Depending 
on how bright such a star is, this saturation can also remove up to 
10\% or more of a single CCD from the survey area.  The Ohio State 
overlays of the POSS survey were used to choose fields with no stars 
as bright as SAO stars \citep{dix81}. 

Thus far we have imaged 134 fields covering a total area of 52 
square degrees.  These fields are listed in Table \ref{nametab}.  Figs.  
\ref{galfig} and \ref{eqfig} show the location of these fields in both 
Galactic and equatorial coordinates. 

\subsection {Instrumentation and Observing Strategy}

As of June 2000, we have had 6 observing runs since October 1998.  The 
observations were taken with 4 different instruments, and three 
different telescopes.  The instruments are as follows: the Burrell 
Schmidt on KPNO, which used a single SITe 2048x4096 CCD with a pixel 
scale of 1.45\arcsec/pixel, and a total area of 1.4 square degrees per 
field; the BTC camera on the CTIO 4m telescope, which consisted of 4 
2048x2048 CCDs with 0.43\arcsec/pixel and a total area of 0.24 square 
degrees per field; the first NOAO Mosaic camera, used on the KPNO 4m 
with 0.27\arcsec/pixel and a total area of 0.38 square degrees per 
field; the second NOAO Mosaic camera used on the CTIO 4m with 
0.25\arcsec/pixel, and a total area of 0.38 square degrees per field.

The most successful observing run was in April 1999 with the BTC 
camera in which we imaged 53 fields.  These data proved to be of 
excellent quality.  This is essential for distinguishing halo 
candidates from foreground dwarfs.  As discussed in 
Sec.~\ref{results}, we were highly successful in identifying halo 
giants from these data.

All four nights in April 1999 were judged photometric through the 
entire night, so, these fields are fully calibrated.  All other 
observing runs had weather related difficulties for at least part of 
the run, and are only partially calibrated.  In subsequent sections we 
describe the April 1999 reduction procedures as a template for the 
general reduction procedures used for this survey.  Along the way we 
describe necessary differences specific to each of the other 
instruments.

Each field's observation consists of a single exposure in each of four 
filters: $C$, $M$, $T2$, and $DDO51$ \citep{can76, gei84}.  The 
respective exposure times for each filter were typically 100, 500, 
120, and 250 seconds.  With limited success we tried to adjust the 
exposure times to match the variations in seeing so that every image 
would have the same photometric sensitivity.  The exposure times, and 
observing conditions, for each field are listed in Table \ref{obstab}.  
With only one exception, the airmass of each exposure is $< 1.8$, and 
most exposures were taken with an airmass $< 1.5$.  The 
seeing ranged from 0.7\arcsec to 3.3\arcsec.

The exceptions to this observing strategy are the 8 fields observed 
with the Burrell Schmidt.  Because it is not possible to achieve the 
necessary photometric depth in a single exposure with this instrument, 
each field was observed at least 4 times under good conditions.  Each 
exposure was 1200-second long in each filter.  We do not list the 
airmass and seeing measurements for individual exposures.  Poor images 
are removed, and photometry from the remaining images are averaged 
prior to being calibrated (see Sec.~\ref{dophot} for a full 
discussion).

For the 4m runs, focus offsets for each filter were determined at the 
beginning of the run.  The telescope was then focussed in a single 
filter at the beginning of each night, and during the night if 
necessary.  The focus was adjusted as the temperature changed 
throughout the night using the standard formula for the 4m telescopes.  
Focus for the Schmidt observations was adjusted throughout the night 
as required.

There is one logistic oddity concerning the BTC chip numbers.  The chip 
numbers referred to in this paper are not the conventional numbers for 
the BTC camera.  It was determined that chips 1 and 2 had been 
exchanged in the readout electronics accidentally during instrument 
setup.  Thus, chip 1 for these observations actually corresponds to 
the traditional chip 2, while chip 2 is actually the traditional chip 
1.  For internal consistency we will refer to each chip by the number 
assigned during this observation run.

\subsection{Photometric Calibration}

It is essential that the data be photometrically calibrated.  
Candidate selection is based on specific color index criteria.  
Furthermore, distance estimates are based on the temperature index and 
magnitude.  It is possible to select candidates based on relative 
positions in the color-color diagrams, however, such candiates will 
need to be calibrated to make them useful.  Photometric standards for 
the Washington filter set \citep{gei90,gei96} were observed on all 
photometric nights.  The strategy was to place a standard field on 
every chip and in every filter at least once during each run to obtain 
zero-point offsets between the chips.  In addition, standard fields 
were observed in all filters with a single chip throughout each night 
to obtain the color term and extinction coefficients.

Some of the fields were observed in non-photometric conditions.  Even 
if a photometric calibration solution is obtained for the instrument 
that took these data, the fields can not be reliably calibrated 
without additional observations.  We have plans to reobserve these 
fields with smaller telescopes, and will publish these data when they 
are fully calibrated.  We have spectroscopically confirmed giants in 
many of these fields, but, of course, do not know their distance.

The only run that was completely photometric was in April 1999 using 
the BTC camera.  We discuss below the determination of this solution.  
We also have standard field observations for the Schmidt, and Mosaic-S 
(Mosaic South on the CTIO 4m; we will refer to the Northern KPNO 
Mosaic camera as Mosaic-N) taken under photometric conditions.  
However, we reserve discussion of these solutions until the full data 
set from these runs is published.
 
The photometric calibration solution for the BTC used three standard 
fields.  Both SA~98 and SA~110 were observed with all four of the BTC 
chips, while NGC~3680 was observed only with chip 4.  The photometric 
standard fields were observed over an airmass range of 1.03 to 1.82, 
encompassing the airmass range of the BTC target fields.  A single 
calibration solution was used for all four nights.  The data were 
examined for variations from night to night, but none were found to a 
precision of $\la 0.005$ magnitudes.

The photometric calibration solution took the form
\begin{mathletters}
\begin{eqnarray}
M  \; = & m + Z_{m} - 0.1424\times X + 0.1538\times (m-t2) \hspace{0.5in}
& \pm \mbox{0.020 (62 stars)}\\
C-M  \; = & Z_{cm} - 0.2052\times X + 1.043\times (c-m) \hspace{0.9in} 
& \pm \mbox{0.023 (63 stars)}\\
M-T2  \; = & Z_{mt2} - 0.09551\times X + 1.158\times (m-t2) \hspace{0.7in}
& \pm \mbox{0.017 (61 stars)}\\
M-51  \; = & m-51^\prime + Z_{m51} + 0.0107\times X + 0.1655\times 
(m-t2) & \pm \mbox{0.018 (65 stars)}
\end{eqnarray}
\label{calibeq}
\end{mathletters}
where $M$, $C$, $T2$ and $51$ correspond to calibrated magnitudes; 
$m$, $c$, $t2$ and $51^\prime$ correspond to instrumental magnitudes; 
$X$ is the airmass; and $Z$ is the zero point.  The color terms and 
the extinction terms were determined using the observations of chip 4; 
no significant differences were found among the color-terms of the 
other chips.  The RMS residuals and the number of stars used in the fit 
are listed to the right of the equations above.  A zero point offset 
was then determined for each of the other chips.  The zero points are 
listed in Table \ref{zerotab}.  There were 5-10 stars available for 
the fit of each chip offset.  The RMS residual ranged from 0.015 to 
0.025.

The final step in our photometric calibration process is to remove 
interstellar reddening.  For this purpose we have used the reddening 
maps of \citet{sch98} to estimate the reddening toward each star.  The 
extinction values for each filter were derived from the relations 
given in \citet{har79}, and \citet{gei84}.  These relations are,
\begin{mathletters}
\begin{eqnarray}
\frac{A_{C}}{E(B-V)} & = & 4.71 \\
\frac{A_{M}}{E(B-V)} & = & 3.64 \\
\frac{A_{51}}{E(B-V)} & = & 3.61 \\
\frac{A_{T2}}{E(B-V)} & = & 2.05 
\end{eqnarray}
\end{mathletters}
assuming $R_{V}=3.3$.  These are high latitude fields, and the 
reddening is quite low by design (see Sec.\ \ref{fields}), so errors 
due to differential reddening are not significant. In addition, the 
$M-51$ index is almost reddening-free.

\section{Photometric Pipeline \label{pipe}}

The processing pipeline for this survey has four main tasks: (1) 
organize the images and associated data files; (2) apply instrumental 
calibration to the images; (3) measure stellar photometry from the 
images; (4) record relevant information about each of these steps.  
Because of the large amount of data, it is also crucial that this 
process is flexible enough to be stopped and started at any point.  
This allows us, for example, to perform some of the processing during 
the observing run, and finish the processing at our home institutions.  
Finally, we require the process to be robust against computer crashes, 
missing header information, and other inevitable mishaps that arise 
during an observing run.  Elements of our approach follow the OGLE 
pipelines.

The essence of our pipeline is to maintain a flag file for each image. 
We chose to keep a separate flag for each chip in the mosaic field, 
although it is possible to maintain a single flag for all the chips 
of a given exposure. This flag file stores information needed by each 
step in the pipeline, namely the location of the image in the computer 
system, and the current reduction state of the image. The former is simply data 
stored in the file itself, while the later is indicated in the flag file 
name. For instance, suppose image spag001.fits has been flat-fielded 
and photometered, while spag002.fits has only been flat-fielded. The 
corresponding flag files might be spag001.phot and spag002.flat. This 
system allows one to glance at a directory of files and know 
immediately the progress of the pipeline.

We have split each step of the process into a separate program or 
script.  These are designed to run continuously as daemon processes.  
Each daemon scans the directory of flag files for images that are 
ready for its step in the process.  When such an image is found, it is 
processed, the flag file is updated, and a log file is appended with 
information about the image relevant to that step.  A separate log is 
maintained by each of the daemons.  Because the pipeline is so 
modular, it is easy to modify or even skip steps without affecting 
other steps in the process.

Fig.\ \ref{flowchart} is a flowchart representation of the reduction 
process.  The shaded region is the instrumental calibration branch.  
These processes are not automated, and must be completed before the 
third daemon can run.  A full description of the instrumental 
calibration steps is discussed in Sec.\ \ref{daemon3}.  In addition, 
photometric calibration and astrometry determinations are not 
currently automated.  We are working on automating the astrometry 
process.

\subsection{Daemon 1: Pipeline Initiation}

The first daemon initiates an image into the pipeline by creating a 
flag file and copying the image to the working area.  Since the flag 
files don't yet exist, this daemon must rely on other clues to 
determine when an image is ready to be processed.  For instance, if 
the images are being processed during the observing run, this daemon 
can scan the acquisition directory, and process any new images that 
appear.  The indications that an image is ready are instrument 
dependent.  One can use the size of the image file, but only if the 
instrument does not pre-allocate the memory for this file at the 
beginning of readout.  To overcome this, once an image file appears 
the daemon can wait for a delay time that is at least as long as the 
readout time.

\subsection{Daemon 2: Organization}

Most often the names generated by the instrument have a prefix and a 
running number, such as spag005.  The second daemon organizes the 
images with convenient names and directories so that images from the 
same field are grouped together.  This daemon renames the file, and 
copies it to the appropriate directory for the field, creating the 
directory if necessary.  A log is created for translating between the 
original name and the new name.  Finally, the new name is placed in 
the flag file so that subsequent daemons know the location of the 
file.

For our survey, the adopted naming convention is
\begin{equation}
l[longitude]b[p/m][latitude]\_[chip]\_[filter]\_[number].
\end{equation}
Bracketed information is variable, and image specific.  We found that 
the nearest degree for longitude and latitude was not quite accurate 
enough to uniquely distinguish different fields.  Hence, the nearest 
tenth of a degree is placed in the name, with no decimal point.  The 
sign of the latitude is indicated by p for the Northern Galactic 
hemisphere and m for the Southern Galactic hemisphere.  The filter is 
given a letter code (M for $M$, G for $DDO51$, I for $T2$, and C for 
$C$), and a reference number.  The reference number is necessary 
because our survey uses several different telescopes, which can have 
different filter sets, and, thus, substantially different color-terms 
from run to run.  Thus, it is necessary to distinguish the $T2$ filter 
of one instrument from the $T2$ filter of another.  Finally, there may 
be more than one image with the same filter in a given field.  A 
running number is appended to the end to distinguish these images.  As 
an example, consider the second exposure of chip 4 from a field at 
$l=100.1$ and $b=-45.5$, with an M filter from the filter set with 
reference code 1.  This would have the name l1001bm455\_4\_M1\_2.fits.

Position, filter, and chip information is read from the header.  In 
those cases where the header information is missing, the daemon stops 
and queries the user to enter the information by hand.  There are also 
cases where images from the same field can be given slightly different 
names.  If the coordinates of different exposures are different by 
just enough to change the nearest tenth of a degree in either Galactic 
latitude or longitude, the names will be different.  An additional 
program was necessary to correct the names of such images. This 
program changes the name, updates the logs to reflect the change, and 
copies the image and associated data files to the correct field 
directory.

The images from each chip of the Mosaic cameras are grouped together 
in a single FITS file when read from the instrument.  Subsequent steps 
in our pipeline work with single image files.  This daemon was 
modified to split the group file into a separate file for each chip.

\subsection{Daemon 3: Instrumental Calibration}
\label{daemon3}

The third daemon applies the instrumental calibration to the images.  
This includes trimming the images, subtracting the bias, subtracting 
the zero-current, applying a flat-field correction, applying an 
illumination correction, and applying a dark sky flat correction.  
Creation of these instrumental calibration images, as described below, is not 
automated (see Fig.\ \ref{flowchart}).

In addition, the instrumental calibration images must be completed 
before this daemon can run.  This presents some limitations in our 
ability to achieve true real-time reduction at the telescope.  The 
most problematic are the dark sky flats, because these come from the 
target images themselves.  We have found that dark sky flats are 
essential for the Mosaic cameras to obtain the best photometry.  We 
are currently testing the reliability of re-using certain calibration 
images, especially the dark sky flats, from previous runs with the 
same instrument.

This daemon was originally written as an IRAF\footnote{IRAF is 
distributed by the National Optical Astronomy Observatories, which are 
operated by the Association of Universities for Research in Astronomy, 
Inc., under cooperative agreement with the National Science 
Foundation.} script to take advantage of the many routines that exist 
to process images.  We have subsequently written a stand-alone program 
to perform these processes.  We achieve an 8-fold speed increase over 
the IRAF script due to optimizing disk access. This speed increase is 
imperative to achieve real-time reduction.

The BTC data were processed with the IRAF script.  In this case the 
overscan region of each line was averaged along the readout direction.  
A first order Legendre polynomial was then fit to these average 
values.  Points which deviated by more than 3 standard deviations were 
eliminated from the fit.  The evaluation of this smooth fit determined 
the bias level subtracted from each line of the image.  The algorithm 
used in all other data sets does not use a smooth fit for the bias 
subtraction.  Instead, the bias for each row was determined separately 
from the average of the overscan region for that row.  This later 
algorithm is also the one used in our stand-alone program.  We found 
this to work better for the Schmidt and Mosaic cameras.  There was no 
significant difference between the two algorithms for BTC images.

Zero-current images are averaged together after being trimmed and 
bias subtracted.  Flat field images are created from dome flats.  
They are trimmed, bias subtracted, zero subtracted, and median 
combined to form a single image for each chip and each filter.

Illumination corrections are made from twilight sky flat images.  The 
twilight images are trimmed, bias-subtracted, zero-subtracted, 
flat-field corrected, and median combined.  A polynomial with order 
between 3 and 5 is then used to fit the large scale structure, 
producing an illumination correction.  The exact order of the fit 
varies depending on instrument, and is chosen as low as possible while 
still reasonably matching the large scale structure.

When combining these instrumental calibration images, they must first 
be scaled.  A significant number of extreme value pixels, from cosmic 
rays or bad pixels, can alter the statistics of the image.  These same 
extreme values also affect the illumination correction fits adversely.  
Therefore, if necessary, we replace these extreme values with more 
appropriate values, such as zero, or the saturation level.

Finally dark sky flats are made from target data.  In the case of the 
April 1999 BTC data, the target images taken on the first two nights 
(April 6 and 7) were used.  The target data are first processed 
normally, including being trimmed, bias-subtracted, zero-subtracted, 
flat-field corrected, and illumination corrected.  The images are then 
median combined.  The median images are examined to be sure all 
objects are adequately removed.  If the objects are not removed, 
clipping is applied during the median combination.  

This dark sky flat also serves to remove the fringe pattern in the 
$T2$ images.  The fringing level is a small fraction of the sky, which 
allows one to represent the fringe as $fringe = \epsilon \times sky$, 
where $\epsilon << 1$ and is approximately constant.  Thus, the net 
count in the image is $sky + fringe = sky (1+\epsilon)$.  The fringe 
pattern has been stable within a given observing run for the 
instruments we have used.  Thus, the normalized dark sky flat contains 
$(1+\epsilon)$.  Division of target images by the dark sky flat leaves 
just $sky$.

The assumption that $\epsilon$ is constant is only approximately true.  
However, because $\epsilon << 1$, small changes in $\epsilon$ lead to 
even smaller errors in the fringe removal.  For example, if $\epsilon$ 
is the dark sky value, while $\epsilon^\prime$ is the target image 
value, the ratio $(1 + \epsilon^\prime)/(1 + \epsilon) = 1 + 
(\epsilon^\prime - \epsilon)/(1 + \epsilon)$.  Thus, if 
$\epsilon^\prime$ is 10\% different from $\epsilon$, the fringe 
correction error is approximately $0.1\epsilon$, which is very small 
since $\epsilon << 1$.

It was found that the BTC data did not require a dark sky correction, 
with the exception of fringe removal.  The magnitude of the dark sky 
correction indicates how well the dome and twilight images flat-field 
the data, as well as provides an indication of flat-field variability.  
The standard deviation of this correction in the central quarter of 
each of the BTC chips was 0.7\%, 0.6\%, 1.4\%, and 0.5\% in $C$, $M$, 
$DDO51$, and $T2$ respectively.  These small corrections indicate the 
chips are remarkably stable, leading to high quality photometry.

Stellar photometry is performed with the DoPHOT \citep{sch93} program 
(see section \ref{dophot}).  The version of DoPHOT used so far in this 
survey requires the image data to be 2 byte integers.  This restricts 
the count values to $\la 32000$.  The saturation level is typically 
larger than this by a factor of two.  To maintain this dynamic range, 
the image values are reduced by a factor of 2 before they are 
converted to 2 byte integers.  The gain is then effectively twice the 
previous value. After division by two the noise in the images was 
still well-sampled.

\subsubsection{Mosaic Cameras}

Both Mosaic instruments received significant benefit from the dark sky 
flat, in addition to fringe removal.  This necessity makes it 
difficult to achieve true real-time processing because many target 
images are needed before a dark sky flat can be constructed.  For 
future runs we will be testing the possibility of using dark sky flats 
made from images taken in previous runs.  The success of this will 
depend on the long-term stability of the instruments, including the 
fringe pattern.

In addition to the basic reduction steps described above, the Mosaic cameras 
require the removal of cross-talk between amplifiers of adjoining 
pairs of chips.  Properly removing this effect requires calibrating 
two chips simultaneously.  This can be done with IRAF, and has been 
implemented in our stand-alone processing code.

\subsubsection{Schmidt Camera}

Prior to being instrumentally calibrated, the Schmidt images require some 
pre-processing. First, the data values are saved by the instrument as 
type short, but actually have the full range of unsigned short by 
using negative values. Thus, the data have to be converted to unsigned 
short and then floating-point to be processed properly.

Second, while the saturation level is constant across the chip, the 
illumination is not.  There is a large flat-field correction at the 
edges compared to the center because of vignetting.  Thus, after 
flat-fielding the images, there is a large variation in the effective 
saturation level.  Current versions of DoPHOT only allow a single 
saturation level for each chip.  Unsaturated, but bright, stars away 
from the center of the chip are pushed above the saturation level by 
the flat-field correction.

If the saturation level is chosen appropriately for the center of the 
chip, these bright stars are flagged as saturated and ignored by 
DoPHOT. These bright stars are needed to define the PSF shape and 
aperture corrections in the outer regions of the chip.  This is 
critical because the PSF is variable.  On the other hand, if the 
saturation is chosen appropriately for the outer regions of the chip, 
saturated stars in the center are accepted as unsaturated resulting in 
inaccurate photometry, and misguided PSF fits.

To overcome this problem, we adjust the saturation level prior to 
flat-fielding the images, but after converting to type float.  All 
values greater than the saturation level are set to an extremely high 
value.  This value must be larger than the saturation level near the 
edges after the flat-field correction is applied.  When the data are 
converted to 2-byte integers for DoPHOT, these extremely high values 
are set to the maximum allowed value to ensure they will be flagged by 
DoPHOT as saturated.

In principle, we could apply this same technique to data from the BTC 
and Mosaic cameras as well.  However, the vignetting for these 
instruments is less severe than the Schmidt's, and the flat-field 
correction varies less across a single chip ($\sim$10\%).  Thus, by 
setting the saturation level to the lowest effective value , we miss 
only a few of the brightest unsaturated stars.

Finally, because of the large field of view of the Schmidt camera, it 
is difficult to obtain good dome flat images. Instead, we relied on 
twilight sky images for the flat-field correction. These are processed 
identically to other twilight sky flats as described above, except we 
do not fit a smooth function to the combined image.

\subsection{Photometry Measurement \label{dophot}}

The final daemon measures stellar photometry from the instrumentally 
calibrated images.  We use the program DoPHOT \citep{sch93} for this 
measurement.  The first step is to create a parameter file for the 
image.  The parameters must include an estimate for the FWHM and the 
sky level.  We wrote a program to estimate these values and output a 
parameter list appropriate for DoPHOT. The sky is estimated as the 
median pixel value.

The FWHM is estimated by finding local maxima in the image, and 
measuring the width at half the peak value.  This is done only in one 
dimension to save computational time.  The median value of these 
widths is used as the estimate.  Because this estimate works in only 
one dimension, some care is needed in restricting the allowed range of 
widths.  Permitting narrow widths, as from cosmic rays or bad pixels, 
produces a FWHM estimate that is too small.  Permitting wide widths 
from a large number of background galaxies, or saturated stars, 
produces a FWHM estimate that is too large.  These limits have to be 
adjusted as the seeing varies.  Thus, the PSF determined by DoPHOT is 
monitored carefully to ensure that the FWHM estimate does not lead the 
program astray.

We set the object detection threshold to 2.5 times the noise standard 
deviation above the background.  The PSF parameters are allowed to 
vary with the third power of position.  The variations are small for 
the BTC, less than 10\% from center to corner within a single chip.

We use an aperture radius of 12 pixels (5.2\arcsec) for determining 
aperture corrections.  This radius was constrained by two factors.  
First, the aperture must be sufficiently large to contain all the flux 
from the star.  This can be determined by plotting the aperture 
photometry as a function of radius.  The aperture photometry will 
converge to a constant value, within the background noise, as the 
radius increases.  Second, the aperture should be as small as possible 
to minimize the noise contribution from the background and 
neighboring stars.  

We have added a routine to DoPHOT for automatically determining and 
applying aperture corrections.  We found that the functional form of 
the aperture correction position variation was the same for the full 
aperture as it was for smaller apertures, provided the smaller 
aperture enclosed the stellar core.  The only difference between the 
smaller aperture function and the full aperture function was a 
constant additive offset.  In effect, the aperture correction did vary 
across a chip, and the shape of this variation could be determined 
from any aperture (within bounds).  Therefore, the functional form of 
the position variation was determined by using a smaller, and hence 
less noisy, aperture.  The aperture correction function determined 
from this smaller aperture was then corrected to a larger aperture 
with a constant offset.

The aperture corrections are determined with a third power positional 
fit.  Only stars with internal errors better than 0.04 magnitudes are 
used in this fit.  To reduce the noise in the fit, the least squares 
determination was made with an aperture radius of 3.6\arcsec.  A 
constant offset was applied to obtain the 5.2\arcsec radius 
correction.  The same procedure is applied to other data sets with 
similar aperture sizes.

The photometry must also be corrected for variable pixel scale.  
DoPHOT works best when the background is truly flat.  Therefore, we 
allow the flat-field process to make the background flat before doing 
photometry.  However, because the pixel scale varies across the field, 
photometric calibration of point sources will be in error 
systematically between the center and edge of the field. The 
correction varies smoothly, and  slowly compared to the PSF size. 
Therefore, we can apply this correction directly to the photometry. 
Care must also be taken to apply the same correction to standard star 
observations. 

Scale variation solutions for both Mosaic instruments are available 
with the MSCRED IRAF reduction package.  We also have a scale variation 
solution for the BTC derived from observations of UNSO standard fields 
(Fischer, P., private communication).  The camera on the Burrell 
Schmidt does not have significant pixel area variation.

Once the final photometry is obtained for all four filters, the stars 
are matched and calibrated.  The star lists of each filter are matched 
to within a 0.5 pixel radius.  All stars with multiple matches within 
this radius are rejected.  The photometric calibration solution is 
then applied.  In order for the color term to be applied, the star 
must be detected in both the $M$ and $T2$ filter.  The stars do not 
have to be detected in $C$ or $DDO51$.  Such stars will not be 
selected as halo giant candidates, although they could be selected 
among the blue halo candidates.

\subsubsection{Schmidt Photometry}
\label{schmidt}

The 8 fields obtained with the Schmidt require a more complicated 
scheme.  Prior to matching between filters, the stellar photometry 
from all of the images in a single filter are matched and averaged.  
The images are often taken on different nights, so the pointing can 
vary from image to image.  Thus, a master image is chosen, and all 
star list coordinates are transformed to this system.

The master also serves as a photometric master.  When averaging 
photometry they must be on the same calibration system.  The 
zero-point can vary from image to image because of clouds, or 
differences in airmass.  A photometric zero-point difference is 
measured and applied to each frame to put them on the same photometric 
system as the master image.  Finally, only stars that were found on 
$\sim 80\%$ of the images are retained.

In addition to averaging the photometry, the coordinates are also 
averaged. The coordinates from each star list are transformed to the 
coordinate system of the master image. The transformation is fit with 
equations of the form
\begin{mathletters}
\begin{eqnarray}
x\prime & = & \delta x + a_{x}x + b_{x}y \\
y\prime & = & \delta y + a_{y}x + b_{y}y
\end{eqnarray}
\end{mathletters}
where $x\prime$ and $y\prime$ are coordinates in the master system, 
and $\delta x$ and $\delta y$ are constants.  With all the coordinates 
in the same system, the coordinates of each star are averaged 
together.  This is necessary to improve positional accuracy because of 
the large pixels of this camera.

\subsection{Astrometry \label{astrometry}}

We determined astrometry solutions using the IRAF package FINDER. The 
USNO-A2 catalog \citep{mon98} was used as a reference for the 
solution.  Unfortunately the position information written to the image 
headers of the BTC files was not consistent.  It varied randomly from 
the true pointing by several arcminutes.  Thus, we were forced to find 
a solution for each chip of each field separately.

The number of reference stars on each chip varied greatly with field 
position, ranging from $<200$ to $>1000$.  Because of this the 
accuracy of the solution varied greatly.  In the low density regions 
the residual RMS from the fit was $\sim 0.5\arcsec$, but was as good 
as $\sim 0.2\arcsec$ in the high density regions.  The position 
information presented in Sec.  \ref{results} should not be considered 
accurate enough for fiber spectroscopy.  However, it is good enough to 
uniquely identify each star. We are working on an automated process 
for determining the astrometric solutions. 

The images taken with the Schmidt camera are often deep enough that 
many astrometric standard stars are saturated.  Thus, to achieve an 
accurate astrometric solution we take several short exposures in $M$ 
of each field.  We average the coordinates of stars detected on 
these short exposures, as we did for the deeper images described in 
Sec.~\ref{schmidt}.  This allows us to find an astrometric solution 
which can be tied to the master frame of the deeper images.  This 
hybrid procedure has produced superb results which have been 
successfully used with fiber-spectroscopes.  Further details of these 
astrometry calculations will be described in a later Paper that 
discusses our fiber spectroscopy.  However, it is worth noting that, 
ultimately, the solution accuracy is limited by the unknown proper 
motions of stars in the USNO-A2 catalog.

\section{Photometry Results \label{results}}

Table \ref{datatab} lists a single star from each of the 53 BTC 
fields.  The complete list of stars with errors $<0.04$ can be 
accessed in the electronic version of this paper, as well as through 
ADC (http://adc.gsfc.nasa.gov/).  Data sets from other observing runs have not yet been 
fully photometrically calibrated because of problematic weather.  As 
these data sets are calibrated we will add them to the ADC database, 
and describe them in subsequent papers in this series.

We found that the photometry from the BTC chip 1 (our notation; 
traditional chip 2) is not as reliable as that from the other three 
chips.  This chip contained a residual flat-field pattern, and the 
photometric calibration data had more scatter.

We have also seen evidence for non-linearity in other data sets with 
this chip.  For example, Fig.~\ref{nonlinfig} shows a star by star 
comparison of photometry measured from two images of the same 
uncrowded field in chip 1.  The two images were taken in December 1998 
two nights apart.  They were processed identically, and photometry was 
measured in the same manner.  They are different in two ways.  Frame 2 
was taken when the sky was $\sim 4$ times brighter than for Frame 1.  
Frame 2 was also taken under better seeing conditions.  They have the 
same exposure time.  The frame with higher counts (due to the sky) 
returns brighter values for the faint magnitudes.  This behavior is 
not seen in the other three chips of these same exposures.  We believe 
the difference is due to non-linear response of the chip.

We have included the stars from chip 1 (traditional chip 2) in the 
plots and tables because we have successfully found halo giants from 
candidates on this chip.  Also,removing these stars from the plots 
does not alter the appearance of the plots, nor decrease the scatter.  
However, the chip number is listed in Table \ref{datatab} so the 
reader can identify these stars if desired.

The DoPHOT internal error distributions for all stars in all fields is 
shown in Fig.  \ref{errfig}.  For reference, the average errors for a 
star at $M=18.5$ are 0.009, 0.008, 0.011, and 0.009 for $C$, $M$, 
$DDO51$, and $T2$ respectively.  Selection for halo candidacy was 
restricted to stars with internal errors better than 0.04.  These 
stars (Table \ref{datatab}) are shown in black, while stars not 
selected are shown in gray.  The error must have been less than this 
cutoff in all filters where the star was detected to have been a 
possible candidate.

The turn-up of stars to higher errors at faint magnitudes is rather 
broad. This is because we have included stars from all 53 fields. The 
distribution from an individual field is much narrower. We achieved 
different depths for each field because of variations in sky 
brightness and seeing. For example, the photometric depth for an 
average $M$ error of 0.04 ranges from $M=20.8$ in field l3263bp386 to 
$M=22.3$ in field l2637bp330.

Fig.\ \ref{mmifig} shows the color-magnitude diagram for stars with 
errors below the cutoff, and for all 53 fields.  We have included 
lines indicating the halo turnoff color index ($M-T2=0.64$) and the 
thick disk turnoff color index ($M-T2=0.80$) (see Paper I for the 
determination of these color indices).

The high quality of these data make two populations readily apparent.  
The thick disk turnoff stars show a concentration with $16 \la M \la 
18.5$ just redward of $M-T2 = 0.80$.  The halo turnoff stars are found 
just redward of $M-T2 = 0.64$.  They are found at all magnitudes, but 
increase in number towards fainter magnitudes.  As discussed in 
Paper~I, stars between these two color indices comprise our halo 
turnoff candidate stars.  Notice that these two populations have sharp 
cutoffs in color index.  This indicates the photometry calibration is 
extremely accurate, so as not to blur these populations when combining 
data from different chips and different fields.

Figs.\ \ref{cmmifig} and \ref{mgmifig} are the color-color diagrams 
for stars with errors below the cutoff, and for all 53 fields.  The 
narrow sequences in these diagrams attest to the high quality of these 
data, as well as the excellence of the relative calibration between 
fields.  These diagrams are used for identifying halo giant 
candidates.  The details of this selection process are outlined in 
Paper~I. Below we provide a brief overview of this process to 
demonstrate our photometric accuracy, and the limitations of 
photometric selection.

$C-M$ is our primary metallicity indicator.  Lower metallicity stars 
of a given effective temperature will have a bluer $C-M$ value.  Halo 
stars are expected to be lower metallicity than disk and thick disk 
stars.  Hence, they are found on the right side of the stellar 
sequence in Fig.\ \ref{cmmifig}.  $M-51$ is sensitive to surface 
gravity.  The dwarfs define a clear sequence that dips to low $M-51$ 
values \citep{gei84, pal94}.  Giant stars, by contrast, are expected 
to follow a different sequence with $M-51$ values that remain near 
zero (see Paper~I).

We have no over-lapping fields to test the relative photometric 
calibration between fields.  However, we can assess the quality from 
the narrowness of the sequence of stars in the color-color diagrams.  
For instance, the RMS deviation of stars about the locus of points in 
Fig.\ \ref{cmmifig} is approximately 0.07.  The scatter for a single 
field is also approximately 0.07.  Hence, the increased scatter due to 
relative calibration errors is at most 0.02.  To provide a visual 
impression of this quality, we have plotted a color-color diagram for 
a single field (Fig.\ \ref{ccd1}).  Each of the BTC chips is marked 
with a different symbol.  None of the chips shows an obvious offset 
with respect to the others.  Hence, we have judged the relative 
photometry between chips and between fields to be excellent.

Halo giant candidates are chosen from among those stars that occupy 
both the low metallicity region in Fig.\ \ref{cmmifig} and the giant 
region of Fig.\ \ref{mgmifig}.  It is evident from Fig.\ \ref{mgmifig} 
that dwarfs far outnumber giants.  In fact, there are not enough 
giants to form a clear sequence.

We have added symbols to Figs.\ \ref{mmifig}, \ref{cmmifig}, and 
\ref{mgmifig} to indicate preliminary results of spectral follow-up on 
a subset of the giant candidates.  The star-symbols indicate stars 
that have been confirmed to be halo giants, square-symbols indicate 
metal-poor subdwarfs, x-symbols indicate dwarfs, and asterisk-symbols 
indicate non-stellar objects.  Our selection of candidates was meant 
to explore the regions of these diagrams and identify the most 
efficient means of selecting giant candidates.

These figures indicate that giant stars are readily found in the 
correct locations, but there is some contamination from dwarfs.  The 
dwarf population arises from two sources.  First, and most 
problematic, are low metallicity subdwarfs.  These are photometrically 
indistinguishable from giants, and we can not eliminate them using 
photometric observations alone.  Second are normal disk dwarfs whose 
photometric errors scatter them into the selection region.  These can 
largely be eliminated with high-precision photometry, and by using 
selections based on both luminosity ($M-51$) and metallicity ($C-M$).

As expected, this contamination is worse for fainter stars.  Fig.\ 
\ref{mgmifig} is divided into three magnitude bins to demonstrate this 
effect.  The confirmed giants occupy a fairly well bounded region 
defined by $-0.02 \la (M-51) \la 0.09$, and $1.1 \la (M-T2) \la 1.8$.  
This region is marked in Fig.\ \ref{mgmifig}.  Note that we have 
observed some stars spectroscopically outside this region in order to 
explore the diagram and optimize future candidate selection.  In fact, 
it was this exploration that helped define this region.  The 
candidates with $(M-51) \la -0.02$ turn out to be predominately 
dwarfs.  The giants not within this bounding box have a fairly high 
metallicity ([Fe/H]$\sim-0.6$), and would not normally pass the $C-M$ 
criteria (see Fig.~\ref{cmmifig}).

The number of observed candidates within the bounding box defined 
above is 40.  Of these, 22 are confirmed giants for a global 
identification efficiency of 55\%.  However, this efficiency is a 
strong function of magnitude, and the overall giant selection 
efficiencies are 75\%, 59\%, and 36\% for $M<17.5$, $17.5<M<18.5$, and 
$18.5<M<19.5$ respectively.

It is more instructive to break these numbers up by subdwarf versus 
dwarf.  The only true failures of the photometric selection are the 
dwarfs, which scatter into the giant region by measurement errors.  
The subdwarfs and non-stellar objects will always be found within this 
region, and cannot be separated from giants by photometry alone.  If 
we restrict the failure count to only dwarfs, the efficiencies, as a 
function of magnitude, are 100\%, 94\%, and 82\% for $M<17.5$, 
$17.5<M<18.5$, and $18.5<M<19.5$ respectively. These are very high 
efficiencies due to the excellent quality of the photometry.

The ratios of giant to subdwarf are 3, 2, and 1 for $M<17.5$, 
$17.5<M<18.5$, and $18.5<M<19.5$ respectively.  This makes going after 
faint, distant giants difficult.  However, we have learned much from 
these first observations, and by adjusting our selection criteria we 
can eliminate many of these subdwarfs.  For example, by increasing the 
lower bound of the selection region (Fig.\ \ref{mgmifig}), most of the 
subdwarfs are eliminated, at the expense of missing several giants.  
At this point it is unclear where the optimal bounds are to maximize 
the number of giants.  Employing such modifications, we expect our 
future efficiency to be even higher.  \citet{mor00b} explores how this 
efficiency is affected by photometric precision, and how to further 
optimize halo candidate selection.

\section{Summary}

We are conducting an imaging survey of high Galactic latitude fields 
to examine the structure of the Galactic halo.  Thus far we have 
observed 134 separate fields covering a total area of 52 square 
degrees.  We have used 3 different telescopes, and 4 different 
instruments.  We present the details of our automated pipeline 
reductions designed to handle this large volume of data.

We present results from our most successful observing run using the 
BTC camera on the CTIO 4m in April 1999.  These data are available in 
the electronic version of this paper, as well as through ADC 
(http://adc.gsfc.nasa.gov/).  All other observing runs occurred in 
partially bad weather, and are thus not fully calibrated.  These data 
will be published and made available on-line in the future as 
calibrating observations are obtained.

We discuss preliminary results in identifying halo giants using the 
Washington filter set.  We achieve an overall giant identification 
efficiency of 66\% for $M<18.5$.  However, most of the non-giant 
candidates are low-metallicity subdwarfs, which cannot be 
distinguished from giants using photometric selection alone.  The 
contamination by dwarfs that scatter into the selection region due to 
photometric errors is only 3\% for $M<18.5$.  This high efficiency is 
only possible with very accurate photometry.  We have also learned 
much from these first observations, and we expect to improve the 
efficiency through refinements in the selection technique.  Details of 
our efficiency and refinements in the selection process will appear in 
a future paper \citep{mor00b}.

\acknowledgments

We would like to thank Doug Geisler, whose efforts with the 
Washington filter set have made this work possible.

This work was supported by NSF grants AST 96-19490 to HLM, AST 
95-28367, AST 96-19632, and AST 98-20608 to MM, and AST 96-19524 to 
EWO.

\clearpage

\figcaption[Dohm-Palmer.fig1.ps]{The locations of the 134 fields in 
Galactic longitude and latitude.  \label{galfig}}

\figcaption[Dohm-Palmer.fig2.ps]{The location of the 134 fields in 
equatorial coordinates.  Also plotted is a grid of Galactic 
coordinates in 30\degr increments.
\label{eqfig}}

\figcaption[Dohm-Palmer.fig3.ps]{A flow chart representation of our 
reduction process.  The instrumental calibration branch is not 
automated, and must be completed before daemon 3 can run.  
\label{flowchart}}

\figcaption[Dohm-Palmer.fig4.ps]{A star by star comparison of 
photometry measured from chip 1 (our notation).  The data comes from a 
previous data set taken in December 1998.  The two images were 
identical except Frame 2 has a higher sky level and better seeing 
conditions.  The photometry clearly deviates at faint magnitude, while 
none of the other three chips show this behavior.  \label{nonlinfig}}

\figcaption[Dohm-Palmer.fig5.ps]{The error distribution of stars for 
all 53 fields.  The black points represent stars with errors below the 
cutoff, 0.04, in all four filters.  The gray points represent stars 
that did not pass this criterion.  \label{errfig}}

\figcaption[Dohm-Palmer.fig6.ps]{The color-magnitude diagram for all 
53 fields.  This plot only includes stars whose errors lie below 0.04 
in all measurements.  Extinction corrections have been applied.  The 
two vertical lines indicate the turnoff color indices for the Halo 
($M-T2=0.64$) and the thick disk ($M-T2=0.8$).  The stars marked with 
special symbols indicate preliminary results from spectroscopic 
observations.  The stars are confirmed Halo giants, the squares are 
metal-poor subdwarfs, the x's are foreground dwarf stars, and the 
asterisks are QSO's and other non-stellar objects.
\label{mmifig}}

\figcaption[Dohm-Palmer.fig7.ps]{A color-color diagram including stars 
from all 53 fields with errors less than 0.04.  Extinction corrections 
have been applied.  Also marked with special symbols are preliminary 
results from spectroscopic observations, as in Fig.  \ref{mmifig}.
\label{cmmifig}}

\figcaption[Dohm-Palmer.fig8.ps]{A color-color diagram including stars 
from all 53 fields with errors less than 0.04.  Extinction corrections 
have been applied.  The plot has been divided into three magnitude 
regions, as indicated in the upper right.  The bounded region is 
where we expect to find giant stars. Also marked with special 
symbols are preliminary results from spectroscopic observations, as in 
Fig.  \ref{mmifig}.
\label{mgmifig}}

\figcaption[Dohm-Palmer.fig9.ps]{ The color-color diagram for field 
l3430bm359.  Each of the BTC chips is marked with a different symbol 
and color.\label{ccd1}}

\renewcommand{\arraystretch}{.6}

\clearpage
\begin{deluxetable}{lrrrrccll}
\tablenum{1}
\tabletypesize{\scriptsize}
\tablecaption{Observed Fields}
\tablewidth{0pt}
\tablehead{\colhead{Name} & \colhead{l} & \colhead{b} & 
\colhead{RA\tablenotemark{a}} & \colhead{DEC\tablenotemark{a}} & 
\colhead{$E(B-V)$} & \colhead{Date\tablenotemark{b}} & 
\colhead{Instrument\tablenotemark{c}} & \colhead{Conditions\tablenotemark{d}}}
\startdata
l0034bp541 &   3.4 &  54.1 & 14:55:00.45 &   6:26:18.2 & 0.034 & 04/08/2000 & Mosaic-S    & \\  
l0037bp613 &   3.7 &  61.3 & 14:32:32.00 &  11:01:32.5 & 0.024 & 04/08/1999 & BTC         & \\
l0055bp488 &   5.5 &  48.8 & 15:14:59.15 &   4:11:14.6 & 0.041 & 04/09/2000 & Mosaic-S    & \\ 
l0064bm190 &   6.4 & -19.0 & 19:17:38.74 & -31:37:27.9 & 0.110 & 04/10/2000 & Mosaic-S    & clouds \\ 
l0111bp440 &  11.1 &  44.0 & 15:39:39.07 &   4:30:43.9 & 0.056 & 04/08/2000 & Mosaic-S    & \\ 
l0118bp517 &  11.8 &  51.7 & 15:14:48.25 &   9:08:53.0 & 0.032 & 04/07/1999 & BTC         & \\
l0171bp467 &  17.1 &  46.7 & 15:39:24.00 &   9:30:59.9 & 0.035 & 04/08/1999 & BTC         & \\
l0182bp470 &  18.2 &  47.0 & 15:39:50.17 &  10:19:09.8 & 0.047 & 04/08/2000 & Mosaic-S    & \\ 
l0420bm414 &  42.0 & -41.4 & 21:35:10.95 & -11:19:20.8 & 0.044 & 10/31/1999 & Mosiac-S    & clouds \\ 
l0481bp813 &  48.1 &  81.3 & 13:29:54.05 &  29:04:31.7 & 0.010 & 02/14/2000 & Mosaic-N    & clouds \\ 
l0592bp857 &  59.2 &  85.7 & 13:09:03.99 &  28:57:44.2 & 0.010 & 02/14/2000 & Mosaic-N    & clouds \\ 
l0645bp855 &  64.5 &  85.5 & 13:09:02.55 &  29:24:54.6 & 0.011 & 02/14/2000 & Mosaic-N    & clouds \\ 
l0658bp855 &  65.8 &  85.5 & 13:08:48.43 &  29:30:12.1 & 0.011 & 02/14/2000 & Mosaic-N    & clouds \\ 
l0820bm610 &  82.0 & -61.0 & 23:37:09.27 &  -4:10:37.4 & 0.036 & 10/31/1999 & Mosiac-S    & clouds \\ 
l0900bm450 &  90.0 & -45.0 & 23:18:55.03 &  11:52:27.7 & 0.043 & 10-11/1998 & B. Schmidt  & clouds \\
l0915bm698 &  91.5 & -69.8 &  0:09:21.24 &  -9:32:21.0 & 0.036 & 10/31/1999 & Mosiac-S    & clouds \\ 
l1000bm500 & 100.0 & -50.0 & 23:52:27.72 &  10:13:39.0 & 0.124 & 10/1998    & B. Schmidt  & clouds \\
l1124bm666 & 112.4 & -66.6 &  0:34:44.62 &  -4:04:12.9 & 0.033 & 10/31/1999 & Mosiac-S    & clouds \\ 
l1200bm400 & 120.0 & -40.0 &  0:41:41.52 &  22:48:58.4 & 0.034 & 11/1998    & B. Schmidt  & clouds \\ 
l1300bm500 & 130.0 & -50.0 &  1:10:01.99 &  12:36:58.8 & 0.041 & 10/1998    & B. Schmidt  & clouds \\
l1350bm400 & 135.0 & -40.0 &  1:31:12.47 &  21:56:16.6 & 0.065 & 10/1998    & B. Schmidt  & clouds \\ 
l1379bm417 & 137.9 & -41.7 &  1:38:44.61 &  19:47:33.7 & 0.046 & 10/30/1999 & Mosiac-S    & \\ 
l1390bm570 & 139.0 & -57.0 &  1:26:14.38 &   4:46:55.3 & 0.022 & 10/31/1999 & Mosiac-S    & clouds \\ 
l1450bp300 & 145.0 &  30.0 &  7:46:02.80 &  70:26:15.4 & 0.028 & 11/1998    & B. Schmidt  & clouds \\ 
l1550bm470 & 155.0 & -47.0 &  2:17:50.81 &  10:25:13.9 & 0.066 & 11/1998    & B. Schmidt  & clouds \\
l1828bp240 & 182.8 &  24.0 &  7:35:14.49 &  36:24:48.4 & 0.048 & 02/14/2000 & Mosaic-N    & clouds \\ 
l1850bm450 & 185.0 & -45.0 &  3:26:09.45 &  -1:35:02.7 & 0.068 & 10/1998    & B. Schmidt  & clouds \\
l1861bm475 & 186.1 & -47.5 &  3:20:06.00 &  -3:42:53.3 & 0.037 & 10/30/1999 & Mosiac-S    & \\ 
l1862bm470 & 186.2 & -47.0 &  3:21:51.60 &  -3:27:58.5 & 0.030 & 10/31/1999 & Mosiac-S    & clouds \\ 
l1886bp302 & 188.6 &  30.2 &  8:11:05.83 &  33:06:47.3 & 0.046 & 02/14/2000 & Mosaic-N    & clouds \\ 
l1994bm354 & 199.4 & -35.4 &  4:22:27.42 &  -5:24:59.6 & 0.045 & 10/31/1999 & Mosiac-S    & clouds \\ 
l2035bp460 & 203.5 &  46.0 &  9:34:01.23 &  25:25:28.3 & 0.020 & 02/14/2000 & Mosaic-N    & clouds \\ 
l2045bp424 & 204.5 &  42.4 &  9:19:29.78 &  23:49:06.8 & 0.034 & 02/14/2000 & Mosaic-N    & clouds \\ 
l2090bp517 & 209.0 &  51.7 & 10:02:09.58 &  23:18:27.8 & 0.033 & 02/14/2000 & Mosaic-N    & clouds \\ 
l2092bm457 & 209.2 & -45.7 &  3:58:04.54 & -16:36:42.7 & 0.029 & 10/30/1999 & Mosiac-S    & \\ 
l2102bp592 & 210.2 &  59.2 & 10:34:44.55 &  24:25:05.3 & 0.019 & 02/14/2000 & Mosaic-N    & clouds \\ 
l2144bp643 & 214.4 &  64.3 & 10:58:27.66 &  23:38:24.1 & 0.018 & 02/14/2000 & Mosaic-N    & clouds \\ 
l2179bp376 & 217.9 &  37.6 &  9:15:35.80 &  12:32:30.5 & 0.025 & 04/09/1999 & BTC         & \\
l2203bm428 & 220.3 & -42.8 &  4:20:45.58 & -23:10:34.0 & 0.031 & 11/01/1999 & Mosiac-S    & \\
l2211bm357 & 221.1 & -35.7 &  4:50:52.50 & -21:38:23.5 & 0.045 & 11/01/1999 & Mosiac-S    & \\
l2223bm477 & 222.3 & -47.7 &  4:01:28.13 & -25:46:11.8 & 0.028 & 11/01/1999 & Mosiac-S    & \\
l2233bp433 & 223.3 &  43.3 &  9:43:52.04 &  11:18:43.6 & 0.019 & 04/09/1999 & BTC         & \\
l2235bp354 & 223.5 &  35.4 &  9:15:43.54 &   7:31:57.8 & 0.054 & 04/09/2000 & Mosaic-S    & clouds \\ 
l2291bm501 & 229.1 & -50.1 &  3:54:12.93 & -30:34:57.6 & 0.011 & 10/30/1999 & Mosiac-S    & \\ 
l2297bp469 & 229.7 &  46.9 & 10:05:28.62 &   9:03:43.6 & 0.032 & 04/08/1999 & BTC         & \\
l2321bp249 & 232.1 &  24.9 &  8:54:35.73 &  -4:11:25.2 & 0.017 & 10/30/1999 & Mosiac-S    & \\ 
l2326bp246 & 232.6 &  24.6 &  8:54:33.91 &  -4:44:04.2 & 0.017 & 04/10/1999 & BTC         & \\
l2328bm252 & 232.8 & -25.2 &  5:48:45.40 & -27:53:36.5 & 0.031 & 11/01/1999 & Mosiac-S    & \\
l2336bm321 & 233.6 & -32.1 &  5:19:23.86 & -30:33:29.0 & 0.016 & 10/30/1999 & Mosiac-S    & \\ 
l2341bp321 & 234.1 &  32.1 &  9:22:34.61 &  -1:42:56.0 & 0.032 & 04/09/1999 & BTC         & \\
l2342bp539 & 234.2 &  53.9 & 10:35:45.99 &  10:15:58.4 & 0.029 & 04/09/1999 & BTC         & \\
l2371bp218 & 237.1 &  21.8 &  8:54:24.75 &  -9:43:42.3 & 0.036 & 04/08/2000 & Mosaic-S    & \\ 
l2373bp416 & 237.3 &  41.6 &  9:59:36.56 &   1:36:45.8 & 0.019 & 04/10/1999 & BTC         & \\
l2376bp354 & 237.6 &  35.4 &  9:40:04.23 &  -2:12:50.6 & 0.027 & 04/09/1999 & BTC         & \\
l2376bp585 & 237.6 &  58.5 & 10:55:37.70 &  11:13:40.1 & 0.018 & 04/08/1999 & BTC         & \\
l2410bp534 & 241.0 &  53.4 & 10:43:29.75 &   6:41:03.5 & 0.034 & 04/08/2000 & Mosaic-S    & \\ 
l2432bp438 & 243.2 &  43.8 & 10:17:10.88 &  -0:28:05.6 & 0.046 & 04/10/1999 & BTC         & \\
l2441bp375 & 244.1 &  37.5 &  9:59:36.18 &  -5:02:22.4 & 0.029 & 04/08/2000 & Mosaic-S    & \\ 
l2449bp632 & 244.9 &  63.2 & 11:19:37.91 &  11:13:06.6 & 0.023 & 04/09/1999 & BTC         & \\
l2468bm314 & 246.8 & -31.4 &  5:34:38.39 & -41:22:10.0 & 0.024 & 10/30/1999 & Mosiac-S    & \\ 
l2481bp300 & 248.1 &  30.0 &  9:45:32.65 & -12:28:15.1 & 0.036 & 04/10/1999 & BTC         & \\
l2522bp266 & 252.2 &  26.6 &  9:45:22.56 & -17:25:43.9 & 0.062 & 04/09/2000 & Mosaic-S    & clouds \\ 
l2524bp528 & 252.4 &  52.8 & 11:00:07.15 &   1:12:47.3 & 0.034 & 04/10/1999 & BTC         & \\
l2525bp597 & 252.5 &  59.7 & 11:19:19.26 &   6:10:51.5 & 0.077 & 04/08/2000 & Mosaic-S    & \\ 
l2574bp405 & 257.4 &  40.5 & 10:37:34.16 & -10:14:58.1 & 0.039 & 04/09/1999 & BTC         & \\
l2586bp483 & 258.6 &  48.3 & 11:00:12.74 &  -4:45:47.5 & 0.035 & 04/10/2000 & Mosaic-S    & clouds\\ 
l2611bp668 & 261.1 &  66.8 & 11:49:42.83 &   9:05:00.9 & 0.024 & 04/09/1999 & BTC         & \\
l2634bp567 & 263.4 &  56.7 & 11:29:37.20 &   0:14:37.5 & 0.023 & 04/09/1999 & BTC         & \\
l2637bp330 & 263.7 &  33.0 & 10:34:41.87 & -19:15:26.9 & 0.055 & 04/08/1999 & BTC         & \\
l2638bp423 & 263.8 &  42.3 & 10:57:32.06 & -11:45:18.1 & 0.033 & 04/08/1999 & BTC         & \\
l2676bp626 & 267.6 &  62.6 & 11:49:32.40 &   4:03:14.0 & 0.026 & 04/08/2000 & Mosaic-S    & \\ 
l2682bp361 & 268.2 &  36.1 & 10:55:02.89 & -18:48:10.1 & 0.037 & 04/10/1999 & BTC         & \\
l2690bp580 & 269.0 &  58.0 & 11:42:36.80 &  -0:15:50.6 & 0.025 & 04/10/1999 & BTC         & \\
l2713bp319 & 271.3 &  31.9 & 10:55:01.87 & -23:43:33.5 & 0.070 & 04/10/2000 & Mosaic-S    & clouds \\ 
l2715bp289 & 271.5 &  28.9 & 10:48:52.59 & -26:23:17.6 & 0.089 & 04/09/2000 & Mosaic-S    & clouds \\ 
l2717bp842 & 271.7 &  84.2 & 12:38:27.98 &  22:07:56.0 & 0.027 & 02/14/2000 & Mosaic-N    & clouds \\ 
l2724bp696 & 272.4 &  69.6 & 12:10:05.92 &   9:13:01.4 & 0.019 & 04/09/1999 & BTC         & \\
l2751bp414 & 275.1 &  41.4 & 11:25:36.60 & -16:47:13.5 & 0.041 & 04/10/1999 & BTC         & \\
l2752bm444 & 275.2 & -44.4 &  3:50:47.96 & -61:52:06.6 & 0.026 & 10/30/1999 & Mosiac-S    & \\ 
l2778bp369 & 277.8 &  36.9 & 11:25:38.67 & -21:44:58.7 & 0.038 & 04/09/2000 & Mosaic-S    & clouds \\ 
l2790bp470 & 279.0 &  47.0 & 11:45:30.94 & -12:47:07.9 & 0.034 & 04/09/1999 & BTC         & \\
l2797bp362 & 279.7 &  36.2 & 11:30:31.12 & -22:59:40.6 & 0.044 & 04/10/1999 & BTC         & \\
l2808bp597 & 280.8 &  59.7 & 12:07:35.85 &  -1:16:28.6 & 0.021 & 04/09/1999 & BTC         & \\
l2815bp423 & 281.5 &  42.3 & 11:45:28.44 & -17:48:36.2 & 0.033 & 04/10/2000 & Mosaic-S    & clouds \\ 
l2822bp413 & 282.2 &  41.3 & 11:46:07.37 & -18:55:38.8 & 0.041 & 04/10/1999 & BTC         & \\
l2843bp365 & 284.3 &  36.5 & 11:46:07.72 & -23:59:51.3 & 0.064 & 04/10/2000 & Mosaic-S    & clouds \\ 
l2881bp416 & 288.1 &  41.6 & 12:04:26.83 & -19:54:52.0 & 0.051 & 04/10/1999 & BTC         & \\
l2897bp367 & 289.7 &  36.7 & 12:04:43.10 & -24:58:06.3 & 0.084 & 04/09/2000 & Mosaic-S    & clouds \\ 
l2900bp487 & 290.0 &  48.7 & 12:16:31.15 & -13:17:34.7 & 0.051 & 04/10/1999 & BTC         & \\
l2914bp438 & 291.4 &  43.8 & 12:16:28.44 & -18:17:14.8 & 0.071 & 04/08/2000 & Mosaic-S    & \\ 
l2924bp617 & 292.4 &  61.7 & 12:31:33.04 &  -0:45:51.8 & 0.027 & 04/08/1999 & BTC         & \\
l2927bp718 & 292.7 &  71.8 & 12:38:33.00 &   9:11:05.2 & 0.020 & 04/08/1999 & BTC         & \\
l2947bp669 & 294.7 &  66.9 & 12:38:31.25 &   4:14:05.8 & 0.032 & 04/09/2000 & Mosaic-S    & clouds\\ 
l2977bp495 & 297.7 &  49.5 & 12:37:29.16 & -13:13:47.8 & 0.049 & 04/10/1999 & BTC         & \\
l2979bp449 & 297.9 &  44.9 & 12:36:28.18 & -17:49:31.6 & 0.037 & 04/08/1999 & BTC         & \\
l3016bp453 & 301.6 &  45.3 & 12:47:30.41 & -17:33:41.6 & 0.042 & 04/09/1999 & BTC         & \\
l3018bp403 & 301.8 &  40.3 & 12:47:41.88 & -22:33:48.6 & 0.079 & 04/10/2000 & Mosaic-S    & clouds \\ 
l3023bp438 & 302.3 &  43.8 & 12:49:30.42 & -19:04:09.7 & 0.066 & 04/10/2000 & Mosaic-S    & clouds \\ 
l3023bp488 & 302.3 &  48.8 & 12:49:43.25 & -14:04:10.6 & 0.054 & 04/10/1999 & BTC         & \\
l3036bp671 & 303.6 &  67.1 & 12:52:28.80 &   4:13:46.7 & 0.038 & 04/09/2000 & Mosaic-S    & \\ 
l3048bp561 & 304.8 &  56.1 & 12:55:38.01 &  -6:45:23.4 & 0.032 & 04/08/2000 & Mosaic-S    & \\ 
l3051bp608 & 305.1 &  60.8 & 12:55:40.21 &  -2:03:14.0 & 0.022 & 04/10/1999 & BTC         & \\
l3051bp611 & 305.1 &  61.1 & 12:55:37.78 &  -1:45:14.7 & 0.022 & 04/08/1999 & BTC         & \\
l3075bp809 & 307.5 &  80.9 & 12:54:28.42 &  18:03:18.8 & 0.028 & 02/14/2000 & Mosaic-N    & clouds \\ 
l3114bp396 & 311.4 &  39.6 & 13:19:43.11 & -22:48:22.5 & 0.116 & 04/10/2000 & Mosaic-S    & clouds \\ 
l3164bp556 & 316.4 &  55.6 & 13:21:52.61 &  -6:28:25.4 & 0.042 & 04/09/2000 & Mosaic-S    & \\ 
l3177bp614 & 317.7 &  61.4 & 13:19:28.46 &  -0:39:54.8 & 0.026 & 04/08/1999 & BTC         & \\
l3201bp350 & 320.1 &  35.0 & 13:53:44.71 & -25:47:09.3 & 0.069 & 04/08/2000 & Mosaic-S    & \\ 
l3220bp398 & 322.0 &  39.8 & 13:53:42.95 & -20:45:16.5 & 0.076 & 04/09/1999 & BTC         & \\
l3236bp441 & 323.6 &  44.1 & 13:52:41.18 & -16:17:59.6 & 0.101 & 04/10/2000 & Mosaic-S    & clouds \\ 
l3251bm766 & 325.1 & -76.6 &  0:25:27.06 & -39:23:19.1 & 0.016 & 10/30/1999 & Mosiac-S    & \\ 
l3258bp548 & 325.8 &  54.8 & 13:43:28.88 &  -5:44:39.8 & 0.026 & 04/10/2000 & Mosaic-S    & clouds \\ 
l3261bp488 & 326.1 &  48.8 & 13:52:43.84 & -11:17:41.9 & 0.057 & 04/08/1999 & BTC         & \\
l3263bp386 & 326.3 &  38.6 & 14:08:51.78 & -20:43:58.7 & 0.079 & 04/07/1999 & BTC         & \\
l3291bm381 & 329.1 & -38.1 & 21:04:14.36 & -65:29:48.0 & 0.029 & 04/09/1999 & BTC         & \\
l3301bp423 & 330.1 &  42.3 & 14:13:47.51 & -16:11:09.1 & 0.080 & 04/08/2000 & Mosaic-S    & \\ 
l3311bm461 & 331.1 & -46.1 & 22:03:07.67 & -60:45:40.4 & 0.038 & 04/09/1999 & BTC         & \\
l3331bp468 & 333.1 &  46.8 & 14:13:33.41 & -11:12:18.4 & 0.064 & 04/08/1999 & BTC         & \\
l3335bp638 & 333.5 &  63.8 & 13:43:28.31 &   4:03:37.6 & 0.026 & 04/09/2000 & Mosaic-S    & \\ 
l3396bp682 & 339.6 &  68.2 & 13:43:21.33 &   9:06:22.9 & 0.026 & 04/09/1999 & BTC         & \\
l3401bp225 & 340.1 &  22.5 & 15:29:34.49 & -28:43:56.6 & 0.440 & 04/09/2000 & Mosaic-S    & \\ 
l3430bm359 & 343.0 & -35.9 & 20:30:00.48 & -54:59:36.9 & 0.048 & 04/10/1999 & BTC         & \\
l3431bp493 & 343.1 &  49.3 & 14:31:26.72 &  -5:36:42.5 & 0.052 & 04/09/2000 & Mosaic-S    & \\ 
l3432bp561 & 343.2 &  56.1 & 14:15:57.75 &  -0:00:58.7 & 0.040 & 04/08/2000 & Mosaic-S    & \\ 
l3440bm483 & 344.0 & -48.3 & 21:51:07.01 & -51:53:20.5 & 0.022 & 10/30/1999 & Mosiac-S    & \\ 
l3443bm434 & 344.3 & -43.4 & 21:19:44.71 & -52:59:45.5 & 0.023 & 04/10/1999 & BTC         & \\
l3469bm493 & 346.9 & -49.3 & 21:53:26.63 & -49:45:52.8 & 0.028 & 04/10/1999 & BTC         & \\
l3479bp533 & 347.9 &  53.3 & 14:31:22.52 &  -0:36:48.7 & 0.046 & 04/10/1999 & BTC         & \\
l3501bp468 & 350.1 &  46.8 & 14:52:25.08 &  -4:41:29.8 & 0.090 & 04/10/2000 & Mosaic-S    & clouds \\ 
l3518bp859 & 351.8 &  85.9 & 13:04:59.79 &  24:23:27.4 & 0.018 & 02/14/2000 & Mosaic-N    & clouds \\ 
l3519bp550 & 351.9 &  55.0 & 14:34:04.08 &   2:12:03.2 & 0.038 & 04/09/2000 & Mosaic-S    & \\ 
l3538bm349 & 353.8 & -34.9 & 20:24:23.60 & -46:10:57.7 & 0.032 & 10/31/1999 & Mosiac-S    & clouds \\ 
l3549bp662 & 354.9 &  66.2 & 14:07:05.51 &  11:17:57.5 & 0.024 & 04/08/1999 & BTC         & \\
l3564bp511 & 356.4 &  51.1 & 14:52:41.12 &   1:16:15.8 & 0.044 & 04/09/1999 & BTC         & \\
\enddata
\tablenotetext{a}{J2000 Coordinates}
\tablenotetext{b}{UT date. The data for each field observed with the Schmidt 
were obtained over 
several nights in October or November of 1998.}
\tablenotetext{c}{BTC indicates the BTC camera mounted on the CTIO 4m; 
Mosaic-N indicates the Northern Mosaic camera mounted on the KPNO 4m; 
Mosaic-S indicates the Southern Mosaic camera mounted on the CTIO 4m; 
B. Shmidt is the Burrell Schmidt at KPNO, owned and operated by Case 
Western Reserve University.}
\tablenotetext{d}{Unless otherwise noted, conditions were photometric}
\label{nametab}
\end{deluxetable}

\clearpage
\begin{deluxetable}{rcccccccccccc}
\rotate
\tablenum{2}
\tabletypesize{\scriptsize}
\tablecaption{Observation summary}
\tablewidth{0pt}
\tablehead{ & \multicolumn{3}{c}{M} & \multicolumn{3}{c}{C} & 
\multicolumn{3}{c}{DDO51} & \multicolumn{3}{c}{T2} \\
\colhead{Name} &
\colhead{Time\tablenotemark{a}} & \colhead{Airmass} & \colhead{FWHM\tablenotemark{b}} & 
\colhead{Time\tablenotemark{a}} & \colhead{Airmass} & \colhead{FWHM\tablenotemark{b}} &
\colhead{Time\tablenotemark{a}} & \colhead{Airmass} & \colhead{FWHM\tablenotemark{b}} & 
\colhead{Time\tablenotemark{a}} & \colhead{Airmass} & \colhead{FWHM\tablenotemark{b}} \\}
\startdata
l0034bp541 & 100 & 1.249 & 1.4 & 500 & 1.257 & 1.4 & 600 & 1.269 & 1.5 & 120 & 1.251 & 1.2 \\  
l0037bp613 & 140 & 1.561 & 1.3 & 700 & 1.504 & 1.3 & 350 & 1.582 & 1.2 & 170 & 1.623 & 0.9 \\ 
l0055bp488 & 100 & 1.447 & 1.0 & 500 & 1.395 & 1.2 & 600 & 1.354 & 1.0 & 120 & 1.427 & 1.0 \\     
l0064bm190 &  50 & 1.028 & 1.2 & 250 & 1.019 & 1.3 & 300 & 1.014 & 1.3 &  60 & 1.024 & 1.1 \\        
l0111bp440 & 100 & 1.233 & 1.4 & 500 & 1.223 & 1.6 & 600 & 1.217 & 1.5 & 120 & 1.228 & 1.1 \\     
l0118bp517 & 100 & 1.372 & 1.0 & 500 & 1.350 & 1.2 & 250 & 1.395 & 1.0 & 120 & 1.385 & 0.9 \\ 
l0171bp467 & 100 & 1.482 & 1.2 & 500 & 1.445 & 1.3 & 250 & 1.496 & 1.1 & 120 & 1.522 & 1.0 \\
l0182bp470 & 100 & 1.776 & 1.4 & 500 & 1.684 & 1.7 & 600 & 1.612 & 1.5 & 120 & 1.739 & 1.3 \\     
l0420bm414 & 120 & 1.124 & 1.5 & 560 & 1.176 & 1.8 & 500 & 1.139 & 1.8 & 140 & 1.142 & 1.8 \\
l0481bp813 & 100 & 1.005 & 0.9 & 500 & 1.003 & 1.2 & 450 & 1.001 & 1.1 & 120 & 1.001 & 1.1 \\
l0592bp857 & 100 & 1.001 & 1.0 & 500 & 1.003 & 1.1 & 450 & 1.006 & 1.2 & 120 & 1.009 & 1.0 \\
l0645bp855 & 100 & 1.036 & 0.8 & 500 & 1.044 & 0.9 & 450 & 1.059 & 0.9 & 120 & 1.070 & 0.8 \\
l0658bp855 & 100 & 1.031 & 0.9 & 500 & 1.024 & 1.0 & 450 & 1.016 & 1.0 & 120 & 1.011 & 0.8 \\
l0820bm610 & 120 & 1.116 & 1.9 & 560 & 1.132 & 2.2 & 500 & 1.122 & 2.5 & 140 & 1.142 & 2.5 \\ 
l0900bm450 & 32x &       &     & 14x &       &     & 11x &       &     &  7x &       &     \\
l0915bm698 & 180 & 1.210 & 1.6 & 840 & 1.298 & 2.1 & 750 & 1.242 & 1.9 & 210 & 1.345 & 2.8 \\  
l1000bm500 &  7x &       &     &  7x &       &     &  6x &       &     &  7x &       &     \\
l1124bm666 & 120 & 1.111 & 1.8 & 560 & 1.118 & 1.7 & 500 & 1.113 & 2.2 & 140 & 1.123 & 2.3 \\  
l1200bm400 &  7x &       &     &  8x &       &     &  9x &       &     &  4x &       &     \\
l1300bm500 & 14x &       &     &  8x &       &     &  7x &       &     &  7x &       &     \\
l1350bm400 & 11x &       &     & 13x &       &     &  9x &       &     &  9x &       &     \\
l1379bm417 & 100 & 1.561 & 1.1 & 500 & 1.581 & 1.2 & 450 & 1.567 & 1.3 & 120 & 1.594 & 1.5 \\         
l1390bm570 & 120 & 1.217 & 1.5 & 560 & 1.218 & 1.7 & 500 & 1.216 & 1.7 & 240 & 1.227 & 2.1 \\  
l1450bp300 & 13x &       &     &  6x &       &     &  9x &       &     &  5x &       &     \\
l1550bm470 &  6x &       &     &  6x &       &     &  8x &       &     &  6x &       &     \\
l1828bp240 & 150 & 1.017 & 3.0 & 750 & 1.012 & 3.0 & 575 & 1.006 & 2.2 & 180 & 1.004 & 1.7 \\
l1850bm450 & 13x &       &     & 15x &       &     &  9x &       &     & 12x &       &     \\
l1861bm475 & 100 & 1.142 & 1.1 & 500 & 1.126 & 1.2 & 450 & 1.135 & 1.2 & 120 & 1.122 & 1.3 \\ 
l1862bm470 & 180 & 1.124 & 2.6 & 840 & 1.118 & 2.0 & 750 & 1.120 & 2.8 & 210 & 1.128 & 3.3 \\  
l1886bp302 & 150 & 1.000 & 1.5 & 750 & 1.002 & 2.1 & 575 & 1.006 & 2.0 & 180 & 1.010 & 1.7 \\
l1994bm354 & 180 & 1.104 & 2.8 & 840 & 1.101 & 3.3 & 750 & 1.101 & 3.3 & 210 & 1.106 & 3.2 \\  
l2035bp460 & 100 & 1.008 & 0.8 & 625 & 1.006 & 0.8 & 580 & 1.007 & 0.8 & 150 & 1.008 & 0.9 \\
l2045bp424 & 150 & 1.045 & 1.4 & 750 & 1.035 & 2.0 & 575 & 1.024 & 2.4 & 180 & 1.018 & 2.2 \\
l2090bp517 & 100 & 1.011 & 0.8 & 500 & 1.011 & 0.8 & 450 & 1.012 & 0.7 & 120 & 1.014 & 0.7 \\
l2092bm457 & 100 & 1.036 & 1.1 & 500 & 1.049 & 1.1 & 450 & 1.041 & 1.2 & 120 & 1.057 & 1.5 \\ 
l2102bp592 & 100 & 1.011 & 0.8 & 500 & 1.009 & 0.9 & 450 & 1.008 & 0.8 & 120 & 1.009 & 0.8 \\
l2144bp643 & 100 & 1.010 & 0.8 & 500 & 1.011 & 0.8 & 450 & 1.013 & 0.8 & 120 & 1.016 & 0.7 \\
l2179bp376 & 100 & 1.361 & 1.0 & 500 & 1.360 & 1.0 & 300 & 1.362 & 1.0 & 120 & 1.366 & 0.9 \\ 
l2203bm428 & 120 & 1.052 & 1.5 & 600 & 1.030 & 1.4 & 540 & 1.042 & 1.3 & 160 & 1.022 & 1.3 \\ 
l2211bm357 & 110 & 1.047 & 1.3 & 550 & 1.028 & 1.4 & 500 & 1.038 & 1.5 & 140 & 1.022 & 1.4 \\ 
l2223bm477 & 120 & 1.294 & 1.7 & 600 & 1.102 & 1.6 & 540 & 1.129 & 1.6 & 160 & 1.085 & 1.8 \\      
l2233bp433 & 100 & 1.350 & 1.0 & 500 & 1.342 & 1.0 & 300 & 1.354 & 0.9 & 120 & 1.363 & 0.8 \\ 
l2235bp354 & 100 & 1.293 & 2.0 & 500 & 1.278 & 2.2 & 600 & 1.269 & 2.0 & 120 & 1.287 & 1.5 \\
l2291bm501 & 100 & 1.000 & 1.1 & 500 & 1.003 & 1.1 & 450 & 1.001 & 1.1 & 120 & 1.006 & 1.3 \\ 
l2297bp469 & 100 & 1.291 & 1.1 & 500 & 1.291 & 1.1 & 250 & 1.291 & 1.0 & 120 & 1.292 & 0.9 \\ 
l2321bp249 & 100 & 1.451 & 1.3 & 500 & 1.364 & 1.4 & 450 & 1.413 & 1.5 & 120 & 1.331 & 1.8 \\ 
l2326bp246 & 100 & 1.109 & 0.8 & 500 & 1.110 & 0.8 & 300 & 1.108 & 0.8 & 120 & 1.107 & 0.8 \\ 
l2328bm252 & 120 & 1.380 & 1.5 & 600 & 1.286 & 1.6 & 540 & 1.339 & 1.6 & 160 & 1.251 & 1.7 \\ 
l2336bm321 & 100 & 1.002 & 1.1 & 500 & 1.000 & 1.4 & 450 & 1.001 & 1.2 & 120 & 1.001 & 1.3 \\ 
l2341bp321 & 100 & 1.143 & 1.3 & 500 & 1.141 & 1.0 & 300 & 1.147 & 1.0 & 120 & 1.150 & 0.8 \\ 
l2342bp539 & 100 & 1.313 & 0.9 & 500 & 1.316 & 1.0 & 300 & 1.312 & 0.9 & 120 & 1.313 & 0.8 \\ 
l2371bp218 & 100 & 1.069 & 1.5 & 500 & 1.090 & 1.7 & 600 & 1.105 & 1.6 & 120 & 1.071 & 1.2 \\
l2373bp416 & 100 & 1.176 & 0.8 & 500 & 1.176 & 0.9 & 300 & 1.176 & 0.8 & 120 & 1.177 & 0.8 \\ 
l2376bp354 & 100 & 1.133 & 0.9 & 500 & 1.134 & 1.0 & 300 & 1.132 & 0.9 & 120 & 1.132 & 0.8 \\ 
l2376bp585 & 100 & 1.334 & 1.3 & 500 & 1.336 & 1.2 & 250 & 1.333 & 1.0 & 120 & 1.332 & 0.8 \\ 
l2410bp534 & 100 & 1.249 & 1.3 & 500 & 1.251 & 1.6 & 600 & 1.257 & 1.5 & 120 & 1.249 & 1.4 \\     
l2432bp438 & 100 & 1.152 & 0.9 & 500 & 1.152 & 1.0 & 300 & 1.153 & 0.9 & 120 & 1.155 & 0.8 \\ 
l2441bp375 & 100 & 1.118 & 1.2 & 500 & 1.110 & 1.2 & 600 & 1.106 & 1.1 & 120 & 1.115 & 1.1 \\     
l2449bp632 & 100 & 1.346 & 0.9 & 500 & 1.338 & 1.0 & 300 & 1.349 & 0.9 & 120 & 1.357 & 0.8 \\ 
l2468bm314 & 100 & 1.020 & 1.1 & 500 & 1.022 & 1.2 & 450 & 1.020 & 1.2 & 120 & 1.024 & 1.2 \\ 
l2481bp300 & 100 & 1.051 & 0.8 & 500 & 1.054 & 0.8 & 300 & 1.050 & 0.8 & 120 & 1.050 & 0.7 \\ 
l2522bp266 & 120 & 1.025 & 1.1 & 750 & 1.028 & 1.8 & 720 & 1.036 & 1.6 & 145 & 1.026 & 1.1 \\      
l2524bp528 & 100 & 1.174 & 0.9 & 500 & 1.171 & 1.0 & 300 & 1.175 & 0.9 & 120 & 1.179 & 0.9 \\ 
l2525bp597 & 100 & 1.251 & 1.3 & 500 & 1.244 & 1.4 & 600 & 1.241 & 1.2 & 120 & 1.248 & 1.1 \\     
l2574bp405 & 100 & 1.075 & 0.8 & 500 & 1.069 & 0.9 & 300 & 1.078 & 0.8 & 120 & 1.083 & 0.8 \\ 
l2586bp483 & 100 & 1.110 & 1.4 & 500 & 1.116 & 1.7 & 600 & 1.126 & 1.5 & 120 & 1.112 & 1.3 \\     
l2611bp668 & 100 & 1.387 & 0.9 & 500 & 1.397 & 1.0 & 300 & 1.371 & 1.0 & 120 & 1.361 & 0.9 \\ 
l2634bp567 & 100 & 1.184 & 0.8 & 500 & 1.189 & 0.9 & 300 & 1.177 & 0.8 & 120 & 1.174 & 0.8 \\ 
l2637bp330 & 100 & 1.019 & 0.9 & 500 & 1.019 & 0.9 & 250 & 1.019 & 0.8 & 120 & 1.020 & 0.7 \\ 
l2638bp423 & 125 & 1.067 & 1.1 & 620 & 1.060 & 1.1 & 310 & 1.071 & 0.9 & 150 & 1.150 & 0.9 \\ 
l2676bp626 & 100 & 1.209 & 1.1 & 500 & 1.213 & 1.2 & 600 & 1.220 & 1.1 & 120 & 1.210 & 1.0 \\     
l2682bp361 & 100 & 1.020 & 0.9 & 500 & 1.020 & 1.0 & 300 & 1.021 & 0.9 & 120 & 1.022 & 0.9 \\ 
l2690bp580 & 100 & 1.209 & 0.9 & 500 & 1.216 & 0.9 & 300 & 1.198 & 0.9 & 120 & 1.192 & 0.9 \\ 
l2713bp319 & 100 & 1.008 & 1.5 & 500 & 1.006 & 1.6 & 600 & 1.007 & 1.5 & 120 & 1.007 & 1.7 \\      
l2715bp289 & 120 & 1.030 & 1.0 & 600 & 1.018 & 1.1 & 720 & 1.010 & 1.2 & 145 & 1.025 & 8.3 \\      
l2717bp842 & 100 & 1.087 & 0.9 & 500 & 1.075 & 1.0 & 450 & 1.060 & 0.8 & 120 & 1.044 & 0.9 \\
l2724bp696 & 100 & 1.402 & 0.9 & 500 & 1.376 & 1.0 & 300 & 1.412 & 1.0 & 120 & 1.434 & 0.8 \\ 
l2751bp414 & 100 & 1.034 & 0.9 & 500 & 1.030 & 1.0 & 300 & 1.035 & 0.9 & 120 & 1.039 & 0.9 \\ 
l2752bm444 & 100 & 1.187 & 1.2 & 500 & 1.179 & 1.2 & 450 & 1.183 & 1.3 & 120 & 1.177 & 1.4 \\ 
l2778bp369 & 110 & 1.016 & 1.3 & 550 & 1.023 & 1.5 & 660 & 1.033 & 1.4 & 132 & 1.018 & 1.1 \\      
l2790bp470 & 100 & 1.328 & 1.1 & 500 & 1.286 & 1.2 & 300 & 1.344 & 1.1 & 120 & 1.377 & 0.9 \\ 
l2797bp362 & 100 & 1.025 & 0.9 & 500 & 1.028 & 0.9 & 300 & 1.020 & 0.8 & 120 & 1.018 & 0.8 \\ 
l2808bp597 & 100 & 1.300 & 1.0 & 500 & 1.312 & 1.0 & 300 & 1.279 & 1.0 & 120 & 1.268 & 0.9 \\ 
l2815bp423 & 100 & 1.038 & 1.2 & 500 & 1.049 & 1.2 & 600 & 1.062 & 1.2 & 120 & 1.042 & 1.1 \\      
l2822bp413 & 100 & 1.092 & 0.8 & 500 & 1.076 & 0.9 & 300 & 1.099 & 0.8 & 120 & 1.112 & 0.8 \\ 
l2843bp365 & 100 & 1.016 & 1.3 & 500 & 1.010 & 1.7 & 600 & 1.007 & 1.5 & 120 & 1.013 & 1.2 \\      
l2881bp416 & 100 & 1.102 & 0.9 & 500 & 1.109 & 1.0 & 300 & 1.089 & 0.9 & 120 & 1.083 & 0.8 \\ 
l2897bp367 & 110 & 1.031 & 1.4 & 550 & 1.020 & 1.7 & 660 & 1.012 & 1.4 & 132 & 1.027 & 1.3 \\      
l2900bp487 & 100 & 1.167 & 0.9 & 500 & 1.144 & 0.9 & 300 & 1.176 & 0.9 & 120 & 1.195 & 0.8 \\ 
l2914bp438 & 100 & 1.037 & 1.0 & 500 & 1.029 & 1.0 & 600 & 1.024 & 1.0 & 120 & 1.034 & 0.9 \\     
l2924bp617 & 100 & 1.147 & 1.0 & 500 & 1.147 & 1.1 & 250 & 1.148 & 1.0 & 120 & 1.149 & 0.9 \\ 
l2927bp718 & 100 & 1.321 & 1.3 & 500 & 1.327 & 1.3 & 250 & 1.314 & 1.0 & 120 & 1.309 & 0.9 \\ 
l2947bp669 & 110 & 1.218 & 2.2 & 550 & 1.228 & 2.0 & 660 & 1.243 & 1.9 & 132 & 1.221 & 1.8 \\     
l2977bp495 & 100 & 1.468 & 1.0 & 500 & 1.410 & 1.0 & 300 & 1.491 & 0.9 & 120 & 1.537 & 0.9 \\ 
l2979bp449 & 100 & 1.027 & 1.0 & 500 & 1.024 & 1.1 & 250 & 1.029 & 1.0 & 120 & 1.032 & 1.0 \\ 
l3016bp453 & 100 & 1.194 & 1.0 & 500 & 1.166 & 1.0 & 300 & 1.205 & 0.9 & 120 & 1.227 & 0.8 \\ 
l3018bp403 & 100 & 1.025 & 1.2 & 500 & 1.017 & 1.4 & 600 & 1.012 & 1.4 & 120 & 1.022 & 1.5 \\      
l3023bp438 & 100 & 1.039 & 1.2 & 500 & 1.052 & 1.5 & 600 & 1.072 & 1.2 & 120 & 1.044 & 1.1 \\      
l3023bp488 & 100 & 1.285 & 0.9 & 500 & 1.299 & 1.0 & 300 & 1.259 & 0.9 & 120 & 1.245 & 0.8 \\ 
l3036bp671 & 100 & 1.213 & 1.5 & 500 & 1.219 & 1.6 & 600 & 1.229 & 2.2 & 120 & 1.215 & 1.3 \\     
l3048bp561 & 100 & 1.121 & 0.9 & 500 & 1.105 & 1.1 & 500 & 1.097 & 1.1 & 120 & 1.116 & 0.8 \\     
l3051bp608 & 100 & 1.738 & 1.1 & 500 & 1.773 & 1.2 & 300 & 1.680 & 1.0 & 120 & 1.645 & 0.9 \\ 
l3051bp611 & 140 & 1.176 & 1.0 & 700 & 1.160 & 1.1 & 350 & 1.183 & 1.1 & 170 & 1.196 & 1.1 \\ 
l3075bp809 & 100 & 1.226 & 1.0 & 500 & 1.202 & 1.0 & 450 & 1.173 & 0.9 & 120 & 1.154 & 0.9 \\
l3114bp396 & 200 & 1.082 & 1.2 & 500 & 1.054 & 1.3 & 600 & 1.040 & 1.2 & 120 & 1.072 & 1.5 \\
l3164bp556 & 100 & 1.131 & 1.3 & 500 & 1.183 & 1.0 & 600 & 1.211 & 1.0 & 120 & 1.159 & 0.9 \\
l3177bp614 & 110 & 1.199 & 1.1 & 540 & 1.205 & 1.2 & 270 & 1.189 & 1.1 & 130 & 1.183 & 1.0 \\ 
l3201bp350 & 100 & 1.003 & 1.0 & 500 & 1.005 & 1.0 & 600 & 1.010 & 1.0 & 120 & 1.004 & 0.9 \\     
l3220bp398 & 100 & 1.224 & 0.9 & 500 & 1.237 & 1.0 & 300 & 1.202 & 1.0 & 120 & 1.190 & 0.8 \\ 
l3236bp441 & 120 & 1.071 & 1.3 & 500 & 1.089 & 1.2 & 600 & 1.109 & 1.2 & 120 & 1.077 & 1.2 \\      
l3251bm766 & 100 & 1.016 & 1.1 & 500 & 1.021 & 1.2 & 450 & 1.027 & 1.2 & 120 & 1.018 & 1.4 \\ 
l3258bp548 & 150 & 1.484 & 1.4 & 750 & 1.397 & 1.4 & 900 & 1.330 & 1.3 & 180 & 1.451 & 1.5 \\
l3261bp488 & 110 & 1.092 & 1.0 & 540 & 1.080 & 1.1 & 270 & 1.097 & 1.0 & 130 & 1.106 & 0.9 \\ 
l3263bp386 & 100 & 1.499 & 1.5 & 500 & 1.384 & 1.5 & 250 & 1.442 & 1.3 & 120 & 1.569 & 1.1 \\ 
l3291bm381 & 100 & 1.521 & 1.2 & 500 & 1.510 & 1.1 & 300 & 1.543 & 1.1 & 120 & 1.556 & 1.1 \\ 
l3301bp423 & 100 & 1.051 & 1.1 & 500 & 1.041 & 1.3 & 600 & 1.035 & 1.0 & 120 & 1.047 & 1.0 \\     
l3311bm461 & 100 & 1.617 & 1.0 & 500 & 1.664 & 1.2 & 300 & 1.601 & 0.9 & 120 & 1.571 & 0.9 \\ 
l3331bp468 & 120 & 1.374 & 1.0 & 580 & 1.394 & 1.1 & 290 & 1.339 & 1.0 & 140 & 1.320 & 0.8 \\ 
l3335bp638 & 100 & 1.217 & 1.7 & 500 & 1.211 & 1.7 & 600 & 1.209 & 1.9 & 120 & 1.215 & 1.2 \\     
l3396bp682 & 100 & 1.458 & 1.2 & 500 & 1.483 & 1.1 & 300 & 1.434 & 1.1 & 120 & 1.421 & 0.8 \\ 
l3401bp225 & 100 & 1.116 & 1.0 & 500 & 1.146 & 1.1 & 600 & 1.178 & 1.1 & 120 & 1.126 & 0.9 \\     
l3430bm359 & 100 & 1.372 & 1.0 & 500 & 1.360 & 1.1 & 300 & 1.395 & 1.0 & 120 & 1.409 & 0.9 \\ 
l3431bp493 & 100 & 1.227 & 0.9 & 500 & 1.265 & 1.1 & 600 & 1.304 & 1.0 & 120 & 1.240 & 0.7 \\     
l3432bp561 & 100 & 1.608 & 1.4 & 500 & 1.459 & 1.4 & 600 & 1.525 & 1.3 & 120 & 1.705 & 1.3 \\
l3440bm483 & 100 & 1.389 & 1.2 & 500 & 1.462 & 1.2 & 450 & 1.418 & 1.4 & 120 & 1.299 & 1.6 \\ 
l3443bm434 & 100 & 1.443 & 1.1 & 500 & 1.486 & 1.1 & 300 & 1.429 & 1.1 & 120 & 1.402 & 1.0 \\ 
l3469bm493 & 100 & 1.459 & 1.1 & 500 & 1.443 & 1.2 & 300 & 1.491 & 1.0 & 120 & 1.511 & 1.0 \\ 
l3479bp533 & 100 & 1.356 & 0.9 & 500 & 1.320 & 1.0 & 300 & 1.370 & 0.9 & 120 & 1.399 & 0.8 \\ 
l3501bp468 & 150 & 1.244 & 1.5 & 750 & 1.295 & 1.5 & 900 & 1.358 & 1.4 & 180 & 1.261 & 1.3 \\      
l3518bp859 & 100 & 1.049 & 0.9 & 500 & 1.060 & 1.0 & 450 & 1.075 & 0.9 & 120 & 1.090 & 1.0 \\
l3519bp550 & 100 & 1.310 & 1.0 & 500 & 1.227 & 1.0 & 600 & 1.211 & 0.9 & 120 & 1.258 & 0.9 \\
l3538bm349 & 120 & 1.136 & 1.4 & 560 & 1.174 & 1.9 & 500 & 1.150 & 1.6 & 140 & 1.193 & 1.7 \\ 
l3549bp662 & 100 & 1.421 & 1.0 & 500 & 1.431 & 1.2 & 250 & 1.406 & 1.1 & 120 & 1.397 & 1.0 \\ 
l3564bp511 & 100 & 1.348 & 1.0 & 500 & 1.315 & 1.1 & 300 & 1.361 & 1.0 & 120 & 1.388 & 1.1 \\ 
\enddata 
\tablenotetext{a}{Time given in seconds.  Fields observed 
with the B. Schmidt telescope have multiple exposures, each of length 
1200 second.  Listed in the table are the number of exposures, 
followed by an ``x''.} 
\tablenotetext{b}{The average FWHM, in 
arcseconds, over all chips.}
\label{obstab}
\end{deluxetable}

\clearpage
\begin{deluxetable}{ccccc}
\tablenum{3}
\tabletypesize{\scriptsize}
\tablecaption{Zero Points for Calibration Solution}
\tablewidth{0pt}
\tablehead{\colhead{chip} & \colhead{$Z_m$} & \colhead{$Z_{cm}$} & 
\colhead{$Z_{mt2}$} & \colhead{$Z_{m51}$}}
\startdata
1 & 25.0311 & -1.2813 & 1.0855 & 2.4780 \\
2 & 25.3346 & -1.0666 & 1.2429 & 2.4498 \\
3 & 25.2730 & -1.0567 & 1.2126 & 2.4592 \\
4 & 25.3753 & -1.1050 & 1.3238 & 2.4716 \\
\enddata
\label{zerotab}
\end{deluxetable}

\clearpage
\begin{deluxetable}{rrrrcrrrrrrrrc}
\rotate
\tablenum{4}
\tabletypesize{\scriptsize}
\tablecaption{Astrometry and Photometry\tablenotemark{a}}
\tablewidth{0pt}
\tablehead{
\colhead{l \tablenotemark{b}} &
\colhead{b \tablenotemark{c}} &
\colhead{RA \tablenotemark{d}} &
\colhead{DEC \tablenotemark{d}} &
\colhead{Chip} &
\colhead{M} &
\colhead{err} &
\colhead{M-T2} &
\colhead{err} &
\colhead{M-51} &
\colhead{err} &
\colhead{C-M} &
\colhead{err} &
\colhead{E(B-V)} \\}
\startdata
  3.7039 &  61.6057 & 14:31:33.11 &  11:12:50.2 & 2 & 18.253 & 0.005 &  1.651 & 0.012 & -0.322 & 0.009 &  1.415 & 0.009 & 0.028 \\
 11.8001 &  51.8498 & 15:14:17.64 &   9:13:44.9 & 2 & 17.074 & 0.005 &  0.872 & 0.009 & -0.054 & 0.012 &  0.628 & 0.006 & 0.031 \\
 17.1003 &  46.9512 & 15:38:30.94 &   9:38:28.9 & 2 & 19.473 & 0.012 &  1.666 & 0.014 & -0.307 & 0.025 &  1.409 & 0.028 & 0.038 \\
217.9004 &  37.4270 &  9:14:57.20 &  12:28:08.6 & 4 & 17.353 & 0.006 &  1.353 & 0.011 & -0.156 & 0.008 &  1.185 & 0.008 & 0.027 \\
223.3000 &  42.9583 &  9:42:37.42 &  11:09:29.0 & 4 & 18.909 & 0.010 &  1.738 & 0.013 & -0.316 & 0.015 &  1.488 & 0.016 & 0.019 \\
229.7000 &  46.6215 & 10:04:30.23 &   8:55:16.5 & 4 & 17.928 & 0.012 &  2.575 & 0.016 & -0.176 & 0.016 &  1.781 & 0.014 & 0.032 \\
232.6007 &  24.2441 &  8:53:21.90 &  -4:55:39.1 & 4 & 16.689 & 0.013 &  0.902 & 0.018 & -0.016 & 0.017 &  0.538 & 0.019 & 0.017 \\
234.1000 &  32.2224 &  9:21:25.81 &  -1:54:28.9 & 4 & 18.608 & 0.004 &  2.899 & 0.010 & -0.056 & 0.008 &  1.782 & 0.009 & 0.041 \\
234.2009 &  53.6885 & 10:35:02.60 &  10:09:06.8 & 4 & 21.080 & 0.017 &  1.215 & 0.026 &  0.015 & 0.035 &  0.849 & 0.032 & 0.030 \\
237.3001 &  41.7252 &  9:58:51.69 &   1:28:40.9 & 4 & 17.498 & 0.007 &  1.223 & 0.009 & -0.155 & 0.011 &  1.149 & 0.009 & 0.022 \\
237.6004 &  35.7009 &  9:39:08.00 &  -2:23:03.8 & 4 & 16.281 & 0.019 &  0.749 & 0.021 & -0.034 & 0.020 &  0.441 & 0.024 & 0.027 \\
237.6140 &  58.3805 &  9:39:12.77 &  -2:23:02.4 & 4 & 18.322 & 0.009 &  0.900 & 0.011 & -0.007 & 0.011 &  0.645 & 0.013 & 0.027 \\
243.2023 &  43.4476 & 10:16:06.12 &  -0:41:41.9 & 4 & 19.628 & 0.010 &  1.261 & 0.016 &  0.023 & 0.015 &  0.707 & 0.013 & 0.049 \\
244.9001 &  63.3978 & 11:20:15.19 &  11:20:40.9 & 1 & 20.422 & 0.013 &  1.732 & 0.024 & -0.449 & 0.039 &  1.394 & 0.029 & 0.024 \\
248.1006 &  30.2162 &  9:45:56.90 & -12:23:00.7 & 1 & 20.122 & 0.010 &  2.869 & 0.012 & -0.112 & 0.022 &  1.842 & 0.027 & 0.034 \\
252.4009 &  52.5560 & 10:59:27.00 &   1:02:07.7 & 4 & 20.625 & 0.012 &  2.369 & 0.015 & -0.257 & 0.028 &  1.649 & 0.026 & 0.030 \\
257.4019 &  40.7193 & 10:38:08.93 & -10:04:58.5 & 1 & 19.440 & 0.010 &  2.148 & 0.016 & -0.310 & 0.019 &  1.696 & 0.020 & 0.038 \\
261.1063 &  66.9456 & 11:50:04.59 &   9:11:54.3 & 1 & 18.154 & 0.009 &  1.025 & 0.010 &  0.016 & 0.013 &  0.609 & 0.013 & 0.024 \\
263.4009 &  57.0064 & 10:33:43.47 & -19:08:45.7 & 2 & 18.307 & 0.020 &  2.155 & 0.028 & -0.268 & 0.028 &  1.702 & 0.024 & 0.058 \\
263.7002 &  32.6650 & 11:30:31.84 &   0:17:06.1 & 1 & 18.490 & 0.009 &  1.707 & 0.017 & -0.351 & 0.017 &  1.472 & 0.012 & 0.024 \\
263.8028 &  41.9794 & 10:35:26.01 & -19:10:06.9 & 1 & 20.859 & 0.018 &  0.608 & 0.033 & -0.108 & 0.036 &  0.202 & 0.025 & 0.062 \\
268.2007 &  36.0021 & 10:55:51.55 & -18:30:03.9 & 1 & 20.409 & 0.014 &  3.087 & 0.015 & -0.121 & 0.030 &  1.864 & 0.034 & 0.039 \\
269.0045 &  58.1317 & 11:42:53.00 &  -0:09:03.6 & 1 & 20.319 & 0.012 &  1.027 & 0.018 & -0.026 & 0.022 & -0.128 & 0.018 & 0.025 \\
272.4004 &  69.3573 & 12:09:38.94 &   9:00:04.1 & 4 & 20.398 & 0.011 &  2.754 & 0.017 & -0.158 & 0.039 &  1.819 & 0.038 & 0.018 \\
275.1008 &  41.6426 & 11:26:03.09 & -16:34:07.7 & 1 & 18.435 & 0.005 &  1.739 & 0.009 & -0.359 & 0.011 &  1.445 & 0.009 & 0.040 \\
279.0020 &  46.8113 & 11:45:14.06 & -12:57:40.9 & 4 & 18.371 & 0.006 &  1.128 & 0.013 & -0.071 & 0.009 &  0.925 & 0.010 & 0.033 \\
279.7000 &  35.9124 & 11:30:10.30 & -23:11:15.8 & 4 & 20.513 & 0.015 &  1.208 & 0.021 & -0.065 & 0.026 &  0.928 & 0.022 & 0.040 \\
280.8007 &  59.9997 & 12:08:00.06 &  -0:59:32.6 & 1 & 19.277 & 0.013 &  0.842 & 0.018 &  0.030 & 0.025 &  0.300 & 0.018 & 0.022 \\
282.2003 &  41.4506 & 11:46:10.30 & -18:53:40.1 & 1 & 19.769 & 0.014 &  1.489 & 0.017 & -0.319 & 0.029 &  1.030 & 0.024 & 0.040 \\
288.1002 &  41.3052 & 12:04:08.60 & -20:12:01.7 & 4 & 20.745 & 0.015 &  2.949 & 0.021 & -0.105 & 0.031 &  1.913 & 0.040 & 0.049 \\
290.0044 &  48.9972 & 12:16:46.83 & -13:00:09.5 & 1 & 18.534 & 0.008 &  0.737 & 0.009 &  0.046 & 0.014 &  0.293 & 0.010 & 0.065 \\
292.4206 &  61.7809 & 12:31:38.51 &  -0:41:10.1 & 1 & 19.544 & 0.010 &  0.728 & 0.016 &  0.039 & 0.018 &  0.234 & 0.014 & 0.024 \\
292.7201 &  72.0990 & 12:32:19.63 &  -0:30:34.8 & 1 & 19.913 & 0.013 &  0.701 & 0.021 & -0.016 & 0.035 &  0.273 & 0.021 & 0.021 \\
297.7721 &  49.6952 & 12:37:44.63 & -13:02:21.5 & 1 & 19.984 & 0.018 &  0.682 & 0.025 &  0.036 & 0.032 &  0.169 & 0.023 & 0.048 \\
297.9800 &  45.0936 & 12:38:15.46 & -13:10:00.7 & 1 & 18.302 & 0.009 &  0.886 & 0.011 &  0.022 & 0.014 &  0.557 & 0.011 & 0.049 \\
301.6746 &  45.5071 & 12:47:44.68 & -17:21:20.4 & 1 & 17.828 & 0.005 &  0.955 & 0.007 & -0.044 & 0.011 &  0.510 & 0.008 & 0.039 \\
302.3018 &  48.9537 & 12:49:43.93 & -13:54:57.6 & 1 & 18.146 & 0.009 &  0.863 & 0.011 & -0.051 & 0.013 &  0.331 & 0.011 & 0.057 \\
305.1494 &  60.5569 & 12:55:48.00 &  -2:17:45.9 & 3 & 19.104 & 0.009 &  2.227 & 0.011 & -0.230 & 0.016 &  1.686 & 0.018 & 0.019 \\
305.1633 &  61.0124 & 12:55:45.85 &  -1:50:26.3 & 3 & 18.382 & 0.008 &  1.637 & 0.009 & -0.320 & 0.011 &  1.436 & 0.011 & 0.020 \\
317.7040 &  61.3817 & 13:19:29.91 &  -0:40:57.4 & 3 & 20.442 & 0.014 &  0.853 & 0.025 & -0.041 & 0.029 &  0.499 & 0.022 & 0.026 \\
322.0002 &  40.0660 & 13:53:21.81 & -20:30:06.2 & 2 & 19.056 & 0.007 &  1.894 & 0.008 & -0.297 & 0.017 &  1.675 & 0.018 & 0.081 \\
326.1006 &  48.6502 & 13:52:57.03 & -11:26:05.2 & 3 & 18.167 & 0.008 &  0.702 & 0.017 &  0.017 & 0.013 &  0.368 & 0.009 & 0.065 \\
326.3003 &  38.4955 & 13:53:38.82 & -11:30:39.6 & 3 & 17.784 & 0.017 &  0.736 & 0.025 &  0.001 & 0.021 &  0.401 & 0.018 & 0.065 \\
329.1007 & -38.4125 & 21:07:04.74 & -65:23:36.5 & 1 & 18.541 & 0.007 &  1.148 & 0.011 & -0.072 & 0.014 &  0.843 & 0.011 & 0.032 \\
331.1019 & -46.5529 & 22:05:34.76 & -60:34:46.9 & 1 & 17.561 & 0.008 &  0.819 & 0.014 &  0.011 & 0.012 &  0.470 & 0.011 & 0.040 \\
333.1022 &  47.0659 & 14:13:04.11 & -10:58:04.1 & 2 & 16.823 & 0.005 &  1.129 & 0.013 & -0.092 & 0.006 &  0.925 & 0.007 & 0.059 \\
339.6040 &  68.2864 & 13:43:05.59 &   9:12:38.5 & 2 & 17.841 & 0.005 &  1.146 & 0.007 & -0.090 & 0.008 &  0.975 & 0.007 & 0.025 \\
343.0137 & -36.2446 & 20:32:20.32 & -54:57:46.8 & 1 & 18.739 & 0.014 &  0.769 & 0.019 &  0.079 & 0.021 &  0.248 & 0.017 & 0.045 \\
344.3014 & -43.7587 & 21:22:04.18 & -52:55:03.0 & 1 & 17.775 & 0.006 &  0.811 & 0.008 &  0.065 & 0.010 &  0.324 & 0.008 & 0.021 \\
346.9024 & -49.4708 & 21:54:27.10 & -49:42:47.1 & 1 & 16.690 & 0.009 &  1.050 & 0.010 & -0.051 & 0.011 &  0.881 & 0.010 & 0.028 \\
347.9004 &  53.3962 & 14:31:08.05 &  -0:32:18.9 & 2 & 19.102 & 0.008 &  1.076 & 0.012 & -0.038 & 0.015 &  0.754 & 0.011 & 0.046 \\
354.9009 &  65.9922 & 14:07:41.91 &  11:09:15.1 & 3 & 19.744 & 0.013 &  3.407 & 0.016 &  0.185 & 0.023 &  1.771 & 0.033 & 0.025 \\
356.4017 &  50.8429 & 14:53:25.44 &   1:05:31.7 & 3 & 19.488 & 0.017 &  2.617 & 0.018 & -0.251 & 0.036 &  1.853 & 0.029 & 0.051 \\
\enddata
\tablenotetext{a}{The complete version of this table is in the 
electronic edition of the Journal. The printed edition contains only 
a sample.}
\tablenotetext{b}{Galactic Longitude}
\tablenotetext{c}{Galactic Latitude}
\tablenotetext{d}{J2000 coordinates}
\label{datatab}
\end{deluxetable}

\end{document}